\documentclass[10pt,aps,prb,twocolumn,showpacs,superscriptaddress]{revtex4-1}
\usepackage{graphicx}
\usepackage{amsmath,amssymb,amsfonts,dsfont}
\usepackage{fbpack} %Copied fbpack.sty from the gradient paper (assuming this unchanged -Bill)
\usepackage{color}
\usepackage[normalem]{ulem} %required for the \sout command to strikeout text
\usepackage{multirow}
\usepackage{array}
\usepackage{upgreek}
\usepackage{mathrsfs}

\newcommand{\temp}{T}

\newcommand{\hQ}{\ham_\mrm Q}
\newcommand{\hF}{\ham_\mrm F}
\newcommand{\hB}{\ham_\mrm B}

\newcommand{\hQF}{\ham_\mrm{QF}}
\newcommand{\hFB}{\ham_\mrm{FB}}

\newcommand{\hMm}{\ham_\mrm M^{(m)}}
\newcommand{\hxodd}{\ham_\xi^\mrm{odd}}
\newcommand{\hxeven}{\ham_\xi^\mrm{even}}
\newcommand{\hamxi}[1]{\ham_\xi^{(#1)}}

\newcommand{\hxi}{\hat\xi}
\newcommand{\wQ}{\w_\mrm Q}

\newcommand{\wn}{\w_n}
\newcommand{\gup}{\gamma_\upa^n}
\newcommand{\gdwn}{\gamma_\dwna^n}

\newcommand{\Tds}{T_2^s}
\newcommand{\Tdst}{T_2^\ast}
\newcommand{\TdM}{T_{2\mrm M}}

\newcommand{\Tde}{T_2^\mrm e}

\newcommand{\ale}{\al^\mrm e}

\newcommand{\mat}{m_\mrm{at}}
\newcommand{\wD}{\w_\mrm{D}}
\newcommand{\vLA}{v_\mrm{LA}}

\newcommand{\wag}{\w_{\al\g}^n}

\newcommand{\wbg}{\w_{\bt\g}^n}
\newcommand{\wg}{\w_\g^n}

\newcommand{\kBT}{\kB\temp}

\newcommand{\wql}{\w_{\mvec q \ld}}

\newcommand{\nF}{n_\mrm F}

\newcommand{\vsq}{\mvec{\sq}}
\newcommand{\vsqab}{\mvec{\sq}_{\al\bt}}

\begin{document}

%\preprint{}

\title{Microscopic models for charge-noise-induced dephasing of solid-state qubits}


\author{F\'elix Beaudoin}
% I have removed the e-mail because I think this is a bad idea since e-mail addresses aren't permanent
% and people will google you if they would like to e-mail.
%\email[]{felix.beaudoin@mail.mcgill.ca}
\affiliation{Department of Physics, McGill University, Montr\'eal, Qu\'ebec, Canada H3A 2T8}
%\thanks{}
%\email[felix.beaudoin@usherbrooke.ca]{Your e-mail address}
%\thanks{}
%\affiliation{D\'epartement de Physique, Universit\'e de Sherbrooke, Sherbrooke, Qu\'ebec, Canada J1K 2R1}
\author{W.~A.~Coish}
\affiliation{Department of Physics, McGill University, Montr\'eal, Qu\'ebec, Canada H3A 2T8}
\affiliation{Canadian Institute for Advanced Research, Toronto, Ontario, M5G 1Z8, Canada}

\date{\today}

\begin{abstract}
	Several experiments have shown qubit coherence decay of the form $\exp[-(t/T_2)^\alpha]$ due to environmental charge-noise fluctuations. We present a microscopic description for temperature dependences of the parameters $T_2$ and $\alpha$.
	Our description is appropriate to qubits in semiconductors interacting with spurious two-level charge fluctuators coupled to a thermal bath. We find distinct power-law dependences of $T_2$ and $\alpha$ on temperature depending on the nature of the interaction of the fluctuators with the associated bath. We consider fluctuator dynamics induced by first- and second-order tunneling with a continuum of delocalized electron states. We also study one- and two-phonon processes for fluctuators in either GaAs or Si. These results can be used to identify dominant charge-dephasing mechanisms and suppress them.
\end{abstract}

% 03.65.Yz -- Decoherence quantum mechanics 
% 72.70.+m -- Noise processes and phenomena in electronic transport in condensed matter
% 74.78.Na -- Mesoscopic and nanoscale systems
% 71.55.-i -- Impurity and defect levels
%\keywords{}

\maketitle

%\section{}
\section{Introduction}

One of the most challenging obstacles to the realization of solid-state quantum computing devices is decoherence caused by charge noise.\cite{dovzhenko2011nonadiabatic,petersson2010quantum,dial2013charge,muhonen2014storing,veldhorst2014addressable,veldhorst2014a}
Charge fluctuations in solid-state devices can arise from several sources, such as Johnson noise from electrical wiring,\cite{muhonen2014storing,nyquist1928thermal,johnson1928thermal} evanescent-wave Johnson noise from metallic gates,\cite{langsjoen2012qubit,poudel2013relaxation} or $1/f$ noise.\cite{paladino2014noise}
The most widely accepted explanation for $1/f$ noise is the presence in the host sample of bistable localized charge states.\cite{dutta1981low} 
Such two-level fluctuators involve tunneling between two spatial configurations with nearly equal potential energy and are routinely observed in amorphous materials.\cite{anderson1972anomalous,black1978interaction,black1981low,agarwal2013polaronic} 
These fluctuators have been observed as spurious resonances in the spectrum of superconducting phase qubits,\cite{simmonds2004decoherence} and have been the subject of an extensive literature in the Josephson qubit community.\cite{lisenfeld2010measuring,grabovskij2012strain,muller2011coherent,dubois2013delocalized,o2015qubit} 
Similar two-level fluctuators consisting of a charge hopping between localized states have been observed in the environment of various other solid-state devices, including lateral gated heterostructures\cite{kurdak1997resistance,jung2004background,pioro2005origin,buizert2008insitu} and self-assembled quantum dots.\cite{kuhlman2013charge,hauck2014locating} Two-level fluctuators have thus been considered an important source of qubit dephasing in several theoretical studies.\cite{faoro2006quantum,sun2010quantum,muller2009relaxation,culcer2009dephasing,ramon2010decoherence,culcer2013dephasing}

Despite the ubiquitousness of two-level charge fluctuators in the solid state, their physical nature can be expected to change from one system to the next. In addition, the microscopic mechanisms causing transitions within pairs of states can hardly be assumed to be universal. For example, the fluctuators can interact with a phonon bath.
\cite{lisenfeld2010measuring,wold2012decoherence}
Alternatively, charge traps near metallic gates or itinerant bands can undergo tunneling.\cite{paladino2002decoherence,desousa2005ohmic,abel2008decoherence,yurkevich2010decoherence}
To minimize the consequent deleterious effects on qubit coherence, it is important to be able to discriminate between different fluctuator baths (e.g., phonons or electrons) from a simple set of measurements.

Any experiment that is designed to measure qubit coherence will typically reveal information about the local environment and may shed light on charge dynamics. Qubit coherence is described by the coherence factor, which empirically often takes the form
\cite{medford2012scaling,dial2013charge}
\begin{equation}
  C(t)=\exp[-(t/T_2)^{\alpha}].\label{eqnSpectro}
\end{equation}
Here, the coherence time, $T_2$, and stretching parameter, $\alpha$, parametrize the decay of qubit coherence. When $\alpha=1$, Eq.~\eqref{eqnSpectro} describes exponential decay, arising from Markovian evolution of the qubit.  For $\alpha\ne 1$, Eq.~\eqref{eqnSpectro} describes a non-Markovian stretched-exponential ($\alpha< 1$) or compressed-exponential ($\alpha> 1$) decay.

The analysis of coherence measurements giving the above empirical form is often based on phenomenological techniques. In the presence of classical Gaussian dephasing noise, $C(t)$ can be written as a simple function of the associated noise spectrum. An analytical form for the noise spectrum is then chosen to best fit the measured coherence factor, $C(t)$.\cite{medford2012scaling,dial2013charge} For example, choosing a $1/f$-like spectrum $S(\nu)\propto 1/\nu^{\beta}$, with $\beta=\alpha-1>0$, exactly yields a coherence factor described by Eq.~\eq{eqnSpectro}.

In this paper, rather than assuming a $1/f$-like spectrum, we begin from a generic microscopic model of fluctuator dynamics. This model results in a coherence factor that closely approximates the compressed-exponential form given in Eq.~\eq{eqnSpectro}. From this model, we find closed-form expressions for the coherence time and stretching parameter, $T_2$ and $\al$. These results allow us to predict a crossover from the non-Markovian to the Markovian regime as temperature $T$ is varied. In addition, we find that different microscopic mechanisms giving rise to fluctuator dynamics typically lead to distinct power-law dependences for $T_2(T)$. In combination with complementary theoretical studies of $T_2(T)$ in the Markovian regime (see, e.g., Refs. \onlinecite{semenov2004phonon,roszak2009phonon,kornich2014phonon}), this will help to better understand and suppress microscopic sources of dephasing.

This paper is divided as follows. In Sec.~\ref{secTwoLevel}, we present the general features of the fluctuator model used throughout the paper. This fluctuator model is used in Sec.~\ref{secCompressed} to show that the qubit coherence factor is well approximated by the compressed exponential form, Eq.~\eqref{eqnSpectro}. In Secs.~\ref{secElectrons} and \ref{secPhonons}, we find analytical expressions for the fluctuator equilibration time and the corresponding noise amplitude for fluctuators coupled to electron or phonon baths, respectively. These expressions are then used in Sec.~\ref{secMicro} to find the temperature dependence of the qubit coherence time $T_2$ and the stretching parameter $\al$ for the microscopic mechanisms considered in Secs.~\ref{secElectrons} and \ref{secPhonons}. We conclude by illustrating an application of this theory to recent experiments.

%%%%%%%%%%%%%%%%%%%%%%%%%%%%%%%%%%%%%%%%%%%%%%%%%%%%%%%%%%%%%%%%%%%%%%%%%%%%%%%%%%%%%%%%%%%%%%%%%%%%%%%%

\section{Two-level fluctuators\label{secTwoLevel}}

\begin{figure}
 \begin{center}
  \includegraphics{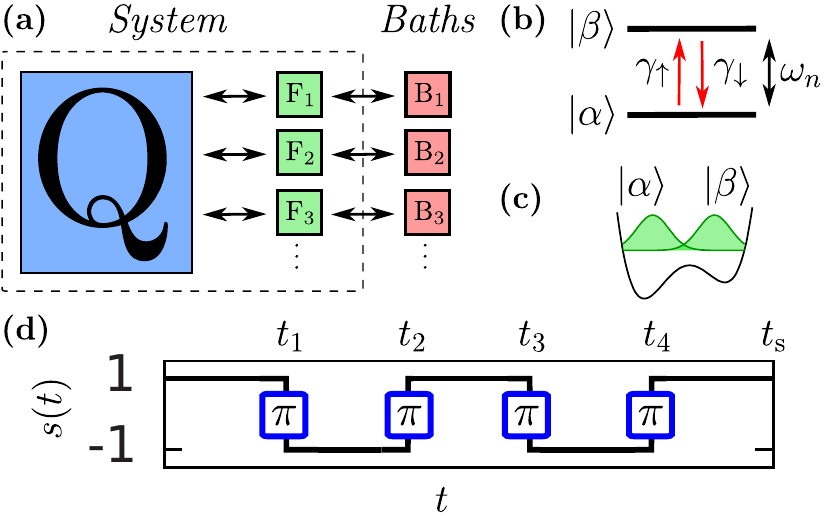}
 \end{center}
 \caption{(Color online)  (a) A qubit (Q) is coupled to an ensemble of independent fluctuators. Each fluctuator $(F_n)$ is itself coupled to an independent bath $(B_n)$. (b)  A two-level fluctuator. Because of the interaction with its bath, the two-level fluctuator is excited at the rate $\gamma_\upa$ and relaxes at the rate $\gamma_\dwna$. (c) Two-level fluctuators can be, e.g., two localized states (represented by the green wave functions) between which a charge can tunnel. (d) The qubit evolves under the influence of sharp control $\pi$-pulses. \label{figQFB}}
\end{figure}

We consider an ensemble of two-level fluctuators coupled to a qubit.  Each fluctuator is itself coupled to an independent thermal bath, allowing equilibration [see Fig.~\ref{figQFB}(a)].  The qubit is subject to a train of fast $\pi$-pulses.  In the toggling frame,\cite{mehring1976high} which accounts for dynamics induced by qubit rotations, the Hamiltonian is then
\begin{equation}
 \ham=\underbrace{\hQ(t)+\sum_n\left[\hF^{n}+\hFB^{n}+\hB^{n}\right]}_{\equiv\ham_0(t)}+\underbrace{\sum_n\hQF^{n}(t)}_{\equiv\hat V},\label{eqnH}
\end{equation}
where
\begin{align}
 \hQ(t)&=\frac12\hbar\wQ s(t)\sz,\qquad\qquad\hF^{n}=\frac12\hbar\w_n\tz_n,\\
 \hQF^{n}(t)&=\frac12\hbar\W_n s(t)\sz\tz_n.\label{eqnHQF0}
\end{align}
Here, we have introduced the Pauli matrices $\sz$ and $\tz_n$ for the qubit and for the $n$-th fluctuator, respectively. The qubit and fluctuator energy splittings are $\hbar\wQ$ and $\hbar\w_n$, respectively, and the qubit-fluctuator couplings are $\hbar\W_n$. The sign function, $s(t)$, alternates between $s(t)=\pm 1$ at times $t_m=t_1, t_2, t_3,\ldots, t_{s-1}$, accounting for a sequence of fast $\pi$-pulses, ending at $t=t_s$ [see Fig.~\ref{figQFB}(d)].  Here we will focus on free-induction decay (no $\pi$-pulse) and Hahn echo (a single $\pi$-pulse),\cite{hahn1950spin} but this notation also allows for a direct extension to other pulse sequences, including, e.g., Carr-Purcell\cite{carr1954effects} or Uhrig dynamical decoupling.\cite{uhrig2007keeping} Retaining only the Ising-like terms $\sim\sz\tz_n$ in the qubit-fluctuator Hamiltonian is justified within a secular approximation, in which the qubit and typical fluctuator splittings are assumed to be large compared to the relevant couplings, $\hbar\omega_Q,\hbar\omega_n \gg \hbar\Omega_n$. The fluctuator-bath interaction $\hFB^{n}$ and bath Hamiltonian $\hB^{n}$ are left unspecified for now. Microscopic forms for these Hamiltonians are considered in Secs.~\ref{secElectrons} and \ref{secPhonons}, where we analyze fluctuator equilibration dynamics for specific physical systems.

To set up a perturbative expansion, we define $\hat V'\equiv\hat V-\mean{\hat V}_F$ and $\ham_0'(t)\equiv\ham_0(t)+\mean{\hat V}_F$, where $\ham_0$ and $\hat V$ are defined in Eq.~\eq{eqnH}. The expectation values $\mean{\cdot}_F$ are taken with respect to the initial state of the fluctuators. We then move to the interaction picture, taking $\hat V'$ as a perturbation (i.e., for a general operator $\hat O$, $\hat O_\mrm I(t)=U_0^\dagger(t)\hat O U_0(t),\;U_0(t) = \exp\left[-i\int_0^t dt' \hat H_0'(t')/\hbar\right]$). We thus have
\begin{align} 
 \hat V'_\mrm I(t)=\frac12\hbar\hat\xi(t)s(t)\sz,\label{eqnHQF}
\end{align}
and we have introduced the noise operator
\begin{align}
 \hat\xi(t)=\sum_n\W_n\left[\hat\tau^z_{n,\mrm I}(t)-\mean{\tz_n}_F\right].	\label{eqnXi}
\end{align}
Our goal is to evaluate the coherence factor parametrized by a pulse sequence $s$,
\begin{equation}\label{eq:CsDefinition}
C^s(t_s) = \left|\langle \hat{S}_+(t_s)\rangle\right|/\left|\langle \hat{S}_+(0)\rangle\right|,
\end{equation}
where $\langle \hat{S}_+(t_s)\rangle=\left[\langle \hat{\sigma}^x(t_s)\rangle+i\langle \hat{\sigma}^y(t_s)\rangle\right]/2$ is the off-diagonal element of the qubit density matrix in the $\hat{\sigma}^z$ eigenbasis. Under quite general conditions, Eq.~\eqref{eq:CsDefinition} can be accurately evaluated using a Magnus expansion.
\cite{blanes2009magnus,magnus1954exponential,maricq1982application}
The leading-order term in the Magnus expansion describes dynamics under the action of the time average of $\hat V'_\mrm I(t)$. This leading-order term will always dominate at sufficiently short time or for sufficiently rapid fluctuations in the noise operator (see Appendix \ref{secMagnus}). Assuming a large number of independent fluctuators, $\hxi(t)$ becomes a source of Gaussian noise due to the central-limit theorem. Conditions for Gaussian noise to dominate over the leading non-Gaussian corrections to the qubit coherence factor are discussed in Appendix~\ref{secMagnus}. We will also assume that the noise is stationary, i.e., that the fluctuators are in a steady state. If, in addition, the initial state of the fluctuators and the qubit is separable, the coherence factor is given by
\begin{align}
 &C^s(t_s)=\eul{-\frac12\int_0^{t_s}dt_1\int_0^{t_s}dt_2s(t_1)s(t_2)g(t_1-t_2)},\label{eqnCts}\\
 &g(t)=\mean{\hxi(t)\hxi(0)},\label{eqnAuto}
\end{align}
where $s\rightarrow\ast$ for free-induction decay and $s\rightarrow\mrm e$ for Hahn echo. In Appendix~\ref{secMagnus}, we consider subleading corrections to the leading-order Magnus expansion and Gaussian approximation. These corrections set limits on the range of validity of Eq.~\eq{eqnCts}.  

In the frequency domain, Eq.~\eq{eqnCts} becomes\cite{martinis2003decoherence,cywinski2008how,uhrig2008exact,desousa2009electron}
\begin{equation}
 C^s(t_s)=\exp\left[-\int_{-\infty}^\infty d\nu\,\frac{S(\nu)}{\nu^2}F^s(\nu t_s)\right], \label{eqnFilter}
\end{equation}
where the noise spectrum $S(\nu)$ and filter function $F^s(\nu t_s)$ are given by
\begin{eqnarray}
S(\nu)&=&\frac{1}{2\pi}\int_{-\infty}^{\infty} dt\,\eul{i\nu t}g(t),\\
F^s(\nu t_s)&=&\frac{\nu^2}{2}\left|\int_0^{t_s} dt s(t)e^{i\nu t}\right|^2.
\end{eqnarray} 
A natural way to describe a compressed-exponential decay [Eq.~\eqref{eqnSpectro}] is to postulate a $1/f$-like noise spectrum,\cite{cywinski2008how,medford2012scaling,dial2013charge,veldhorst2014addressable}
\begin{equation}
S(\nu) = \frac{A}{|\nu|^\beta},\label{eqn1overfNoise}
\end{equation}
with a general exponent $\beta$. Such a spectrum can also be justified from noise-spectroscopy measurements.\cite{bylander2011noise,yuge2011measurement} Inserting the $1/f$-like spectrum, Eq.~\eqref{eqn1overfNoise}, into Eq.~\eqref{eqnFilter} leads directly to a compressed-exponential decay [Eq.~\eqref{eqnSpectro}] with stretching parameter $\alpha$ and coherence time $T_2^s$ given by
\begin{eqnarray}
\alpha &=& 1+\beta,\label{eq:alpha1overf}\\
T_2^s &=& \left(2A\int_0^\infty dx\frac{F^s(x)}{x^{\alpha+1}}\right)^{-1/\alpha}.	\label{eqnT21surf}
\end{eqnarray} 
$T_2^s$ exists when the integral in Eq.~\eq{eqnT21surf} converges, i.e., when $\al<2$ for free-induction decay (since $F^\ast(x)\propto x^2$ for $x\rightarrow0$) and when $\al<4$ for Hahn echo (since $F^\mrm e(x)\propto x^4$ when $x\rightarrow0$). One consequence of Eq.~\eqref{eqn1overfNoise} is that the stretching parameter $\alpha$ depends only on the noise spectrum through the exponent $\beta$ [Eq.~\eqref{eq:alpha1overf}], not on the pulse sequence $s$.  This procedure provides a satisfying and useful relationship between the stretching parameter $\alpha$, coherence time $T_2^s$, and pulse sequence $s$.  However, ultimately, Eq.~\eqref{eqn1overfNoise} amounts to a (non-unique) reparametrization of the observed compressed-exponential decay and does not necessarily provide additional insight into the relevant physical processes or further predictive power.  An alternative approach, which we take here, is to directly evaluate fluctuator dynamics from plausible microscopic interactions.

Equation \eq{eqnCts} shows that for a given pulse sequence, $C^s(t_s)$ is entirely determined by the autocorrelation function $g(t)$ of the fluctuator-induced noise. To evaluate this autocorrelation function, we consider the regime where the fluctuator dynamics are described by a Markovian master equation. The evolution of a fluctuator is Markovian when the fluctuator equilibrates with its local bath on a time scale $\tau_n$ that is long compared to the bath correlation time $\tau_\mrm{cB}^n$. Typically, $\tau_\mrm{cB}^n$ is set by the inverse bandwidth of bath excitations. When $\tau_n\gg\tau_\mrm{cB}^n\,\forall\,n$, the evolution of the fluctuators is described by a Lindblad-form master equation. Assuming, as illustrated in Fig.~\ref{figQFB}(a), that each fluctuator is coupled to an independent bath, the reduced density matrix, $\hat\rho_n$, for fluctuator $n$ evolves according to 
\begin{align}
 \dot{\hat\rho}_n(t)&=\liouv_n\hat\rho_n(t),\\
 \liouv_n\cdot&=-\frac{i}{\hbar}[\hF^n,\cdot]+\gup\dissip[\tpl_n]\cdot+\gdwn\dissip[\tm_n]\cdot,\label{eqnLiouv}
\end{align}
where $\dissip[\hat X]\hat O=\hat X\hat O\hat X^\dagger-\frac12(\hat X^\dagger \hat X \hat O+\hat O\hat X^\dagger \hat X)$ and where $\g^n_\upa$ and $\g^n_\dwna$ are the excitation and relaxation rates for fluctuator $n$ [see Fig.~\ref{figQFB}(b)]. In the above equation, and throughout this paper, the centerdot (``$\cdot$'') represents an arbitrary operator upon which the relevant superoperator is applied. Using Eq.~\eq{eqnLiouv}, it is then straightforward to evaluate $g(t)$ with the usual multitime averaging formula.\cite{gardiner2000quantum} Under the stationary-noise assumption, the autocorrelation function of the resulting noise becomes that of a mixture of independent Ornstein-Uhlenbeck processes,\cite{uhlenbeck1930theory,wang1945theory}
\begin{equation}
 g(t_1-t_2)=\sum_n \Delta\xi^2_n\eul{-|t_1-t_2|/\tau_n}\label{eqnOU}.
\end{equation}
Here, $\Delta\xi_n$ is the amplitude of the noise induced by fluctuator $n$ and $\tau_n$ is the associated equilibration time.  These parameters are related directly to the excitation (relaxation) rates $\gamma_{\uparrow(\downarrow)}^n$ and couplings $\Omega_n$ through 
\begin{align}
 \Delta\xi_n^2&=\W_n^2\frac{4\g_\upa^n\g_\dwna^n}{[\g_\upa^n+\g_\dwna^n]^2},\label{eqnDxind}\\
 1/\tau_n&=\g_\upa^n+\g_\dwna^n.\label{eqnTau}
\end{align}
We note that Eqs.~\eq{eqnOU} to \eq{eqnTau} would be unchanged if a pure dephasing term $\propto\dissip[\tz_n]\cdot$ were added to Eq.~\eq{eqnLiouv}.

As is well known, a mixture of Ornstein-Uhlenbeck processes, Eq.~\eqref{eqnOU}, can approximate $1/f$ noise with an appropriately chosen distribution of amplitudes and equilibration times.\cite{surdin1939fluctuations,bernamont1937fluctuations,mcwhorter19571,dutta1981low,paladino2014noise} It is not, however, generally necessary to approximate a $1/f$-like noise spectrum [Eq.~\eqref{eqn1overfNoise}] to find a coherence factor $C^s(t_s)$ that approximates a compressed-exponential decay.  As we illustrate numerically below, even a Lorentzian noise spectrum associated with a single equilibration time $\tau=\tau_n$ results in an approximate compressed-exponential decay over a wide parameter range. 

%%%%%%%%%%%%%%%%%%%%%%%%%%%%%%%%%%%%%%%%%%%%%%%%%%%%%%%%%%%%%%%%%%%%%%%%%%%%%%%%%%%%%%%%%%%%%%%%%%%%%%%%

\section{Functional form of the coherence factor \label{secCompressed}}

Substituting the noise autocorrelation function [Eq.~\eq{eqnOU}] into the coherence factor [Eq.~\eq{eqnCts}] with the function $s(t)$ for either free-induction decay ($s\rightarrow\ast$) or Hahn echo ($s\rightarrow\mrm e$) gives the closed-form expressions,\cite{klauder1962spectral,desousa2009electron}
\begin{align}
 C^s(t_s)&=\exp\left[-f^s(t_s)\right],\label{eqnCsts}\\
 f^s(t_s)&=\frac{t_s}{\TdM}-\sum_n\D\xi_n^2\tau_n^2\,h^s(t_s/\tau_n),\label{eqnfs}\\
 1/\TdM&=\sum_n\D\xi_n^2\tau_n,\label{eqnTdM}
\end{align}
where
\begin{align}
 &h^\ast(x)=1-\eul{-x},\label{eqnhst}\\
 &h^\mrm e(x)=\eul{-x}-4\eul{-x/2}+3.\label{eqnhe}
\end{align}

\begin{figure}
 \begin{center}
  \includegraphics{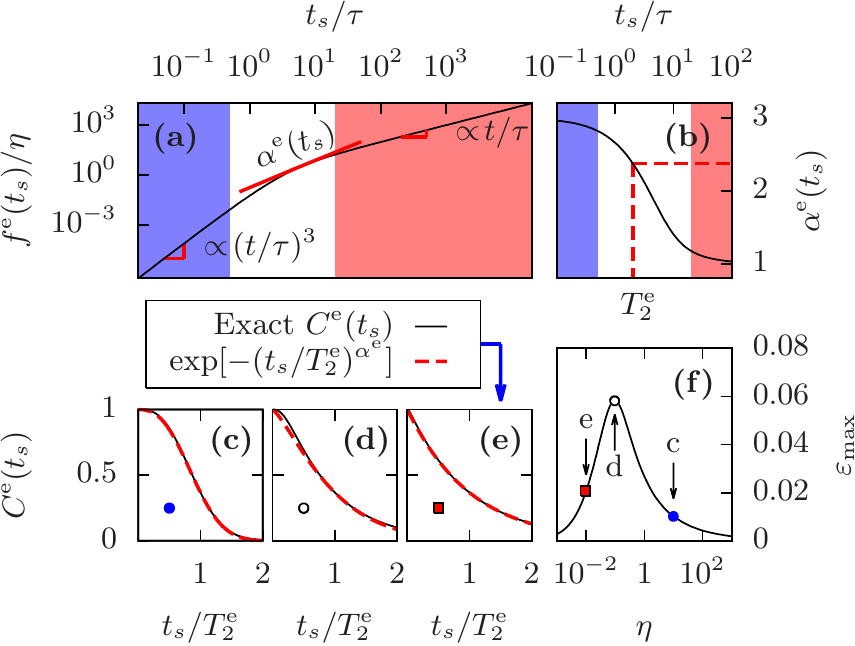}
 \end{center}
 \caption{(Color online) Approximate compressed-exponential form of $C^\mrm e(t_s)$ when $\tau_n\equiv\tau\,\forall\,n$. (a) The slope of $f^\mrm e(t_s)$ in log-log scale gives $\al^\mrm e(t_s)$ [see Eq.~\eq{eqnAlpha}]. (b) We approximate $C^\mrm e(t_s)\simeq\exp[-(t_s/\Tde)^{\al^\mrm e}]$ by taking $\al^\mrm e\equiv\al^\mrm e(\Tde)$. When $\ale(\Tde)\simeq3$ (in the slow-noise limit, where $\Tde/\tau \ll 1$ is in the blue area), decay is faster than exponential and $\Tde$ is given by Eq.~\eq{eqnTdNMe}. When $\ale(\Tde)\simeq1$ (in the fast-noise limit, where $\Tde/\tau \gg 1$ is in the red area), the decay is purely exponential and $\Tde \simeq T_\mrm{2M}$ is given by Eq.~\eq{eqnTdM}. (c-e) Comparison of the exact (solid black line) and compressed-exponential (dashed red line) forms of $C^\mrm e(t_s)$. (c) $\eta=10$, (d) $\eta=0.1$, (e) $\eta=0.01$. (f) Maximum error made by taking $C^\mrm e(t_s)\simeq\exp[-(t_s/T_2^\mrm e)^{\al^\mrm e}]$ with $\al^\mrm e\equiv\al^\mrm e(T_2^s)$, Eq.~\eqref{eq:epsmax}. Dots correspond to (c-e). \label{figCrossover}}
\end{figure}

We define $T^s_2$ to be the $1/e$ decay time of $C^s(t)$ through
\begin{equation}
 f^s(T_2^s)=1.	\label{eqnDefT2s}
\end{equation}
The form of $C^s(t_s)$ as given in Eq.~\eqref{eqnCsts} does not generally describe a pure compressed-exponential decay, $\sim \exp\left[-\left(t/T_2\right)^\alpha\right]$.  However, we will show that Eq.~\eqref{eqnCsts} can approximate a compressed exponential over a wide parameter range.  We therefore define a time-dependent stretching parameter $\al^s(t_s)$ such that, instantaneously,  $f^s(t_s)=(t_s/T_2^s)^{\al^s(t_s)}$ and introduce a typical value $\al^s$ of the stretching parameter at the $1/e$ decay time,
\begin{equation}
 \al^s(t_s)=\frac{d\log f^s(t_s)}{d\log(t_s/T_2^s)},\quad \alpha^s\equiv \alpha^s(T_2^s).\label{eqnAlpha}
\end{equation}
The functions $f^\mrm e(t_s)$ and $\al^\mrm e(t_s)$ are shown in Figs.~\ref{figCrossover}(a) and \ref{figCrossover}(b) assuming $\tau_n\equiv\tau\,\forall\,n$. The coherence factor can then be replaced by $C^s(t_s)\simeq\exp[-(t_s/T_2^s)^{\al^s}]$ with small corrections when $\al^s(t_s)$ varies slowly for $t_s$ in the vicinity of $T_2^s$.

The coherence factor behaves very differently in either the ``slow-noise" or ``fast-noise" regime.  These two regimes are determined by the ratio of the correlation time $\tau_c$ [the decay time of the noise autocorrelation function $g(t)$] to the coherence time, $T_2^s$.  We define the correlation time $\tau_c$ through\cite{fick1990quantum}
\begin{equation}
 \tau_c\equiv\frac{\int_0^\infty dt g(t)t}{\int_0^\infty dt g(t)}=\frac{\sum_n \D\xi_n^2\tau_n^2}{\sum_n \D\xi_n^2\tau_n},\label{eqnTauc}
\end{equation}
where the second equality follows directly from Eq.~\eqref{eqnOU}.

The slow-noise regime is given by $T_2^s< \tau_c$.  In this regime, $g(t)$ is slowly-varying in Eq.~\eqref{eqnCts} over the time scale of interest ($\sim T_2^s$).  Expanding $g(t)$ around $t=0$ and keeping the leading nontrivial correction in Eq.~\eq{eqnCts} then gives the compressed-exponential form in Eq.~\eq{eqnSpectro}, with $\alpha=\alpha^s$ and $T_2=T_2^s$ for decoupling sequence $s$, consistent with known results for Gaussian spectral diffusion due to classical noise,\cite{klauder1962spectral}
\begin{align}
 \al^\ast=2,\quad1/T_2^\ast&=\textstyle{\left(\frac12\sum_n\D\xi_n^2\right)^{\frac12}},\hspace{10mm} (T_2^\ast \ll \tau_c),\label{eqnTdNMst}\\
 \al^\mrm e=3,\quad 1/T_2^\mrm e&=\textstyle{\left(\frac{1}{12}\sum_n\D\xi_n^2/\tau_n\right)^{\frac13}},\quad (T_2^\mathrm{e} \ll \tau_c).\label{eqnTdNMe}
\end{align}

In the opposite (fast-noise) regime, $T_2^s\gtrsim \tau_c$, we evaluate $T_2^s$ and $\alpha^s$ from Eqs.~\eqref{eqnDefT2s} and \eqref{eqnAlpha}.  Neglecting exponentially small corrections in $T_2^s/\tau_c\gtrsim 1$, we find the coherence times
\begin{eqnarray}
T_2^\ast &= & \frac{(1+\sum_n\Delta\xi_n^2\tau_n^2)}{\sum_n\Delta\xi_n^2\tau_n},\hspace{18.5mm} (T_2^\ast \gtrsim \tau_c),\label{eqnT2StarFast}\\
T_2^\mrm e &= & \frac{(1+3\sum_n\Delta\xi_n^2\tau_n^2)}{\sum_n\Delta\xi_n^2\tau_n},\hspace{16.5mm} (T_2^\mrm e \gtrsim \tau_c),\label{eqnT2eFast}
\end{eqnarray}
and stretching parameters
\begin{eqnarray}
\alpha^s & = &  1+\beta^s,\label{eqnBetaDef}\\
\beta^\ast & = & \sum_n \Delta\xi_n^2\tau_n^2,	\hspace{28mm} (T_2^\ast \gtrsim \tau_c), \label{eqnBetaStar}\\
\beta^\mrm e & = & 3\sum_n \Delta\xi_n^2\tau_n^2,\hspace{26mm} (T_2^\mrm e \gtrsim \tau_c).\label{eqnBetaE}
\end{eqnarray}
In contrast with the result from an assumed $1/f$-like spectrum in Sec.~\ref{secTwoLevel}, here the stretching parameter $\alpha^s$ is sensitive to the pulse sequence $s$. In fact, the parameters $\beta^s$ for echo and free-induction decay are related by a universal factor of three in the fast-noise regime, $\beta^e\simeq 3\beta^*$.  

Equations \eqref{eqnTdNMst} to \eqref{eqnBetaE} provide a complete analytical description of both the coherence time $T_2^s$ and form of decay (through $\alpha^s$)  in either the slow-noise or fast-noise regime.  This description can be related to a microscopic model of fluctuator dynamics through the noise amplitudes $\Delta\xi_n^2$ and equilibration times $\tau_n$.  In particular, $T_2^s$ and $\alpha^s$ will inherit temperature dependences associated with the fluctuator excitation (relaxation) rates $\gamma^n_{\uparrow (\downarrow)}$ through Eqs.~\eqref{eqnDxind} and \eqref{eqnTau}.  In the rest of this paper, we will evaluate these temperature dependences for physically relevant microscopic mechanisms and connect fluctuator dynamics to qubit coherence through Eqs.~\eqref{eqnTdNMst} to \eqref{eqnBetaE}.  Since the qubit coherence time $T_2^s$ and noise correlation time $\tau_c$ typically have distinct temperature dependences, tuning the bath temperature will typically induce a transition between the slow-noise $(T_2^s<\tau_c)$ and fast-noise $(T_2^s\gtrsim\tau_c)$ regimes.  

To describe the transition from the slow-noise to the fast-noise regime, it is useful to define a dimensionless parameter that controls a Markov approximation:
\begin{equation}
\eta\equiv\frac{\tau_c}{\TdM}=\sum_n\D\xi_n^2\tau_n^2.	\label{eqneta}
\end{equation}
When $\eta\ll1$ (the fast-noise limit), a Markov approximation gives exponential decay ($\al^s=1$), with $T_2^\ast \simeq T_2^\mrm e \simeq\TdM$.  In the opposite (slow-noise) limit, $\eta\to\infty$, we recover the results of Eqs.~\eqref{eqnTdNMst} and \eqref{eqnTdNMe}.

While the coherence factor exhibits a simple form in either the slow-noise or fast-noise limit, it is less clear how to simply describe the decay in the intermediate regime $\eta\sim 1$.  It is, however,  straightforward to numerically verify the assumed compressed-exponential form $\left\{C^s(t_s)= \exp{\left[-(t/T_2^s)^{\alpha^s}\right]}\right\}$. To simplify the analysis, we assume a single equilibration time for all fluctuators, $\tau_n\equiv\tau\,\forall\,n$, corresponding to a pure Lorentzian noise spectrum $S(\nu)$. In this case, Eq.~\eq{eqnfs} reduces to
\begin{equation}
 f^s(t_s)=\eta\left[\frac{t_s}\tau-h^s(t_s/\tau)\right].\label{eqnfsSingleTau}
\end{equation}
In Figs.~\ref{figCrossover}(c-e), we compare $C^\mrm e(t_s)$ with $\exp[-(t_s/T_2^\mrm e)^{\al^\mrm e}]$ for a fixed correlation time $\tau$ and a range of $\eta$. For a given value of $\eta$, the maximum error made in replacing $C^s(t_s)$ by the compressed-exponential form is 
\begin{align}\label{eq:epsmax}
 \veps_\mrm{max}\equiv\max_{t_s\in[0,\infty[}\{|C^s(t_s)-\exp[-(t_s/T_2^s)^{\al^s}]|\}.
\end{align}
In Fig.~\ref{figCrossover}(f), we plot $\veps_\mrm{max}$ as a function of $\eta$. Dots in Fig.~\ref{figCrossover}(f) indicate the three values of $\eta$ corresponding to Figs.~\ref{figCrossover}(c-e). The error, $\veps_\mrm{max}$, is maximized for $\eta\simeq0.1$, the value taken for Fig.~\ref{figCrossover}(d). Even in this worst case, the difference between the exact and compressed-exponential forms of $C^\mrm e(t_s)$ is small ($\veps_\mrm{max}\simeq0.06$). Thus, while the microscopic analysis presented here leads, in general, to a complex functional form [Eq.~\eqref{eqnCsts}], this functional form will likely be indistinguishable from a compressed exponential in many experiments.

%%%%%%%%%%%%%%%%%%%%%%%%%%%%%%%%%%%%%%%%%%%%%%%%%%%%%%%%%%%%%%%%%%%%%%%%%%%%%%%%%%%%%%%%%%%%%%%%%%%%%%%%

\section{Electron baths \label{secElectrons}}

\begin{figure}
 \begin{center}
  \includegraphics{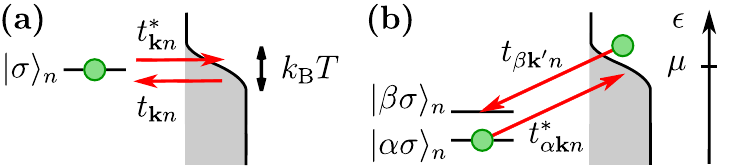}
 \end{center}
 \caption{(Color online) Tunneling processes between localized electron states and a continuum of delocalized states. (a) Direct tunneling. This first-order process is only allowed if $|\e_n-\mu|\lesssim\kBT$, where $\e_n$ is the energy of the localized state for fluctuator $n$ and $\mu$ is the chemical potential of the electron reservoir. (b) Cotunneling between pairs of localized states forming a fluctuator $n$. This second-order process occurs if $|\e_{\al n}-\e_{\bt n}|\lesssim\kBT$. \label{figTunnel}}
\end{figure}

In this section, we consider charge fluctuators described by Anderson impurities. These impurities can equilibrate through tunnel coupling to a continuum of delocalized electronic states in a reservoir (the bath).  The electron reservoirs are held in thermal equilibrium with occupation described by a Fermi-Dirac distribution $\nF(\e)=1/\{\exp[(\e-\mu)/\kBT]+1\}$ at a common temperature $T$ and chemical potential $\mu$. As illustrated in Fig.~\ref{figTunnel}, we consider both first-order (direct tunneling) and second-order (cotunneling) processes. Qubit decoherence due to fluctuators tunnel-coupled to an electron reservoir has been considered previously in, e.g., Refs.~\onlinecite{abel2008decoherence,grishin2007low,paladino2002decoherence,desousa2005ohmic,yurkevich2010decoherence}.

\subsection{Direct tunneling	\label{secDirect}}

In the first-order process (direct tunneling), we assume that each impurity $n$ is coupled to an independent bath through a Fano-Anderson model.\cite{mahan1990many} We then have
\begin{align}
 &\hF^n=\sum_{\s}\eps_n\dd_{n\s}\od_{n\s},	\qquad	\hB^n=\sum_{\mvec k\s}\eps_{\mvec k}\cd_{\mvec kn\s}\lc_{\mvec kn\s},\label{eqnFreeElectrons}\\
 &\hFB^n=\sum_{\mvec k\s}\left(t_{\mvec kn}^\ast\cd_{\mvec kn\s}\od_{n\s}+t_{\mvec k n}\dd_{n\s}\lc_{\mvec kn\s}\right). \label{eqnHFano}
\end{align}
For each fluctuator $n$, we have introduced $\od^{(\dagger)}_{n\s}$ and $\lc^{(\dagger)}_{\mvec kn\s}$, the annihilation (creation) operators for the localized and delocalized states, respectively. The corresponding eigenenergies are $\eps_n$ and $\eps_{\mvec k}$. The spin index is $\s\,\in\,\{\upa,\dwna\}$ and $t_{\mvec k n}$ is the amplitude for tunneling between the impurity and the continuum. Assuming strong Coulomb blockade for each impurity (due, e.g., to a large on-site charging energy), we restrict to the space of singly-occupied ($\ket\al\equiv\ket{\s}_n$) and empty ($\ket\bt\equiv\ket{0}_n$) states. Thus, each impurity $n$ is a two-level fluctuator with splitting $\hbar\wn=\e_n-\mu$.  Each impurity can couple to the qubit through the Coulomb interaction.\cite{gamble2012two,ramon2010decoherence} Under these assumptions, Eqs.~\eqref{eqnFreeElectrons} and \eqref{eqnHFano} correspond to the physical model of Eq.~\eq{eqnH}.

In direct tunneling, we find the excitation ($\gup$) and relaxation ($\gdwn$) rates of a given fluctuator using Fermi's golden rule,
\begin{align}
 &\g_{\al\rightarrow\bt}^n=\frac{2\pi}{\hbar}\sum_{if}\rho(i)|\,\!_n\bra{\bt f}\hFB^{n}\ket{\al\,i}_n|^2\dt(\hbar\wn+E_f-E_i),\label{eqnFermi}
\end{align}
where $\alpha$ and $\beta$ are collective indices (including, e.g., both spin and orbital degrees of freedom) labeling the initial and final states of the fluctuator, $i$ and $f$ label the initial and final states of the bath with energies $E_i$ and $E_f$, respectively, and $\rho(i)$ is the probability for the bath to be initially in state $i$. In thermal equilibrium, this probability distribution is given by the Fermi-Dirac distribution. We take the continuum limit $\sum_{\mvec k}\rightarrow\int d\e D_\mrm{el}(\e)$ when summing over the initial and final bath states $i$ and $f$. 
Using Eqs.~\eq{eqnDxind} and \eq{eqnTau}, we calculate the noise amplitudes $\Delta\xi_n$ and the fluctuator equilibration rates $1/\tau_n$ from $\gup$ and $\gdwn$. Summing the rates of the transitions from the reservoir to the degenerate eigenstates $\ket{\!\!\upa}$ and $\ket{\!\!\dwna}$ then gives
\begin{align}
 \D\xi_n^2&=\frac{8\W_n^2\eul{\hbar\wn/\kBT}}{(2+\eul{\hbar\wn/\kBT})^2},\label{eqnXiTunnel}\\
 \frac1{\tau_n}&=\frac{2\pi}{\hbar} D_\mrm{el}(\e_n)|t_n(\e_n)|^2\frac{2+\eul{\hbar\wn/\kBT}}{1+\eul{\hbar\wn/\kBT}},\label{eqnTauTunnel}
\end{align}
where $t_n(\eps_n)$ is the tunneling amplitude $t_{\mvec kn}$ in the continuum limit. Equation \eq{eqnXiTunnel} implies that the fluctuators are frozen out and have exponentially small contribution to qubit dephasing when $\hbar|\wn|=|\e_n-\mu|>\kB T$, as expected from Fig.~\ref{figTunnel}. In the opposite (high-temperature) limit, $\kBT>\hbar|\wn|$, we have $\D\xi_n^2\simeq\frac89\W_n^2$, giving a maximal contribution to qubit dephasing. In this high-temperature limit, Eq.~\eq{eqnTauTunnel} also gives an equilibration rate that is approximately constant with temperature.

\subsection{Cotunneling}

In the second-order tunneling process (cotunneling), we consider the case where two localized states with energies $\e_{n\al}$ and $\e_{n\bt}$ are coupled to the same electron reservoir $n$. We now have
\begin{align}
 \hF^n=\sum_{l\s}\e_{nl}\od^\dagger_{ln\s}\od_{ln\s},\;\;\;\hat V^n=\sum_{l\mvec k\s}\left(t^\ast_{l\mvec kn}\cd_{\mvec kn\s}\od_{ln\s}+\hc\right),	\label{eqnVCotunnel}
\end{align}
where $l\in\{\al,\bt\}$. In this case, $\hB^n$ is again given by Eq.~\eqref{eqnFreeElectrons}. When $\mu-\e_{ln}>\kBT\;\forall\;l\in\{\al,\bt\}$, direct tunneling is forbidden. However, the cotunneling process illustrated in Fig.~\ref{figTunnel} can still occur if $\e_{\bt n}-\e_{\al n}<\kBT$. Each fluctuator $n$ is then described by a pair of localized states coupled to the same bath with fluctuator energy splitting $\hbar\wn=\e_{\bt n}-\e_{\al n}$. The fluctuator-bath Hamiltonian corresponding to the (second-order) cotunneling process is obtained using the Schrieffer-Wolff expansion. To leading order in $\hat V_n$, the effective Hamiltonian for this process can be written as
\begin{align}
 \hFB^n=\frac12\commut{\left(\frac{1}{\mrm L_0^n}\hat V^n\right)}{\hat V^n}, \label{eqnSW}
\end{align}
where $\mrm L_0^n\cdot=[\hF^n+\hB^n,\cdot]$. Using this fluctuator-bath Hamiltonian, we evaluate excitation and relaxation rates using Fermi's golden rule, Eq.~\eq{eqnFermi}. As written, Eq.~\eqref{eqnSW} contains formal divergences (zero denominators) corresponding to resonant cotunneling processes.  These contributions can be systematically regularized,\cite{koenig1997cotunneling} leading to exponentially small corrections in the limit  $\mu-\e_{ln}>\kBT\;\forall\;l$, which we assume here. Neglecting resonant cotunneling in this limit, from the inelastic cotunneling rates,\cite{qassemi2009stationary} we then find
\begin{align}
 \frac{1}{\tau_n}&\simeq\frac{1}{\pi}\left|\frac{\hbar \G_n}{\mu-\overline{\e}_n}\right|^2\wn\coth\left(\frac{\hbar\wn}{2\kBT}\right),\label{eqnRateCotunnel}\\
 \G_n&=2\pi D_\mrm{el}(\mu)|t_{\al n}(\mu)||t_{\bt n}(\mu)|/\hbar,\\
 \D\xi_n^2&=\W_n^2\,\mrm{sech}^2\left(\hbar\wn/2\kB \temp\right).\label{eqnsech}
\end{align}
Here, we have introduced $\overline{\e}_n\equiv(\e_{\al n}+\e_{\bt n})/{2}$. Eq.~\eq{eqnRateCotunnel} is valid up to corrections of order $\sim\hbar\w_n/(\mu-\overline{\e}_n)$. The difference between $\D\xi_n^2$ given in Eq.~\eq{eqnsech} and that given in Eq.~\eq{eqnXiTunnel} for direct tunneling arises from spin degeneracy.\cite{beenakker1991theory,gurvitz1996microscopic} As in the case discussed below Eq.~\eq{eqnXiTunnel}, Eq.~\eq{eqnsech} implies that $\D\xi_n^2$ decays exponentially for $\hbar\w_n>\kBT$. However, from Eq.~\eq{eqnRateCotunnel}, for $\kBT>\hbar\wn$ the equilibration rate $1/\tau_n$ now increases linearly with $T$.

Table~\ref{tabTunnel} summarizes the distinct temperature dependences obtained for $1/\tau_n$ due to the two processes discussed in this section. These will be useful in Sec.~\ref{secMicro}, when we evaluate the temperature dependences of $T_2^s$ and $\al^s$.
\begin{table}
 \begin{center}
  \begin{tabular}{lc}
   \hline
   \hline
      & $1/\tau_n$ \\
   \hline
   Direct tunneling	& $\propto 1$\\
   Cotunneling		& $\propto T$\\
   \hline   
   \hline
  \end{tabular}
 \end{center}
 \caption{Temperature dependence of the equilibration rates $1/\tau_n$ for electronic baths when $\hbar\wn<\kBT$ in the case of first-order (direct) tunneling and second-order cotunneling. In both cases, $1/\tau_n$ is independent of the fluctuator splitting $\wn$. \label{tabTunnel}}
\end{table}

%%%%%%%%%%%%%%%%%%%%%%%%%%%%%%%%%%%%%%%%%%%%%%%%%%%%%%%%%%%%%%%%%%%%%%%%%%%%%%%%%%%%%%%%%%%%%%%%%%%%%%%%

\section{Phonon baths \label{secPhonons}}

In this section, we evaluate the amplitude $\D\xi_n^2$ and equilibration time $\tau_n$ for fluctuators coupled to independent phonon baths. For all processes considered in this section, $\D\xi_n^2$ is given simply by Eq.~\eq{eqnsech}, valid in the absence of spin degeneracy. This expression can be derived from Eq.~\eq{eqnDxind} simply by assuming detailed balance between $\gup$ and $\gdwn$. To evaluate $\tau_n$, we will consider one-phonon direct, and two-phonon sum and Raman processes, as indicated schematically in Figs.~\ref{figPhonons}(a-c).

Each fluctuator consists of two impurity states $\ket{\al}$ and $\ket{\bt}$. These could be, e.g., two localized states in a double well, as illustrated in Fig.~\ref{figQFB}(c), or the ground and excited states of a single donor impurity. The energy splitting between states $\ket \al$ and $\ket \bt$ for fluctuator $n$ is $\hbar\w_n\equiv\e_{\bt n}-\e_{\al n}$. Thus, the fluctuator and bath Hamiltonians for fluctuator $n$ are
\begin{align}
 \hF^n=\sum_{l\s}\e_{nl}\dd_{ln\s}\od_{ln\s},\;\;\;	\hB^n=\sum_{\mvec q\ld}\hbar\w_{\mvec q\ld}\ad_{\mvec qn\ld}\oa_{\mvec qn\ld},	\label{eqnHFandBphonons}
\end{align}
where $\oa^{(\dagger)}_{\mvec qn\ld}$ annihilates (creates) a phonon with wave vector $\mvec q$ in branch $\ld$ of the phonon bath $n$. We work within the regime of validity of the envelope-function approximation for the impurity. We also assume acoustic phonons with a linearized dispersion. We will focus on two materials: GaAs and silicon. For either material, ignoring anharmonic corrections, the fluctuator-bath interaction is then given by
\begin{equation}
 \hFB^n=\sum_{\s \mvec q \ld\chi}\sum_{l\neq l'}A_{\mvec q\ld\chi} S_{\chi,ll'}^n(\mvec q)\dd_{nl\s}\od_{nl'\s}(\oa_{\mvec qn\ld}+\ad_{-\mvec qn\ld}),\label{eqnElectronPhonon}
\end{equation}
where $l,l'\in\{\al,\bt\}$. In Eq.~\eqref{eqnElectronPhonon}, we have introduced the electron-phonon coupling strength
\begin{equation}
 A_{\mvec q\ld\chi}=A_{\mvec q\ld\chi}^\mrm d-iA_{\mvec q\ld}^\mrm p,
\end{equation}
where d and p label the deformation and piezoelectric contributions, respectively. The form of these contributions is given in Appendix~\ref{secElPhon} in terms of material parameters. In Eq.~\eq{eqnElectronPhonon}, we have also introduced the form factor
\begin{align}\label{eq:FormFactor}
  S_{\chi,ll'}^n(\vq)=\int d\vecr |\al_\chi\vf_\chi(\vecr)|^2 F_{\chi l}^{n\ast}(\vecr)F_{\chi l'}^{n}(\vecr)\eul{i\vq\cdot\vecr},
\end{align}
where $\vphi_\chi(\mvec r)$ is the Bloch amplitude with wave vector $\mvec k_\chi$ corresponding to the degenerate conduction-band minimum (valley) $\chi$, and $F_{\chi l}^n(\vecr)$ is the corresponding envelope function for impurity state $l$ of fluctuator $n$. $\al_\chi$ is the coefficient for valley $\chi$ appearing in the wave function $\sum_\chi\al_\chi F_{\chi l}^n(\mvec r)\vphi_\chi(\mvec r)$ of impurity state $l$.

The coupling between pairs of impurity states is suppressed if they are separated by more than the impurity size, $\ell_\mrm{imp}$, describing the extent of the envelope $F_{\chi l}(\mathbf{r})$ [see Eq.~\eqref{eqnsq}, below]. Here, we assume $\ell_\mrm{imp}$ satisfies
\begin{equation}
 \ell_\mrm{imp}<\hbar v_\ld/\kBT\;\forall\;\ld,	\label{eqnab}
\end{equation}
where $v_\ld$ is the phase velocity of branch $\ld$. Under the above condition, the typical phonon wavelength $2\pi/q_\mrm{th}\sim hv_\ld/\kBT$ is much longer than the spacing between coupled impurity states. The form factor $S_{\chi,ll'}^n(\vq)$ defined in Eq.~\eq{eqnElectronPhonon} can then be approximated in the small-$q$ (long-wavelength) limit,
\begin{align}
 S_{\chi,ll'}^n(\vq)&\simeq i|\al_\chi|^2\vq\cdot\vsq^{\chi n}_{ll'},	\label{eqntmullp}\\
 \vsq^{\chi n}_{ll'}&=\int d\vecr\,\vecr|\vf_\chi(\vecr)|^2F_{\chi l}^{n\ast}(\mvec r) F_{\chi l'}^n(\mvec r)	\label{eqnsq},
\end{align}
where $\vsq_{ll'}^{\chi n}$ is the transition dipole matrix element between states $l$ and $l'$.
To obtain Eq.~\eq{eqntmullp}, we have used the first non-vanishing term of a Taylor expansion around $\vq=0$. This amounts to neglecting phonon-bottleneck effects,\cite{benisty1995reduced} which suppress the contribution from short-wavelength (high-energy) phonons having a typical wavelength on the order of the impurity spacing. For $v_\ld=3070\;\mrm{m/s}$ (the smallest phase velocity among all the relevant branches in GaAs and silicon) and $T=100$~mK, Eq.~\eq{eqnab} implies that these bottleneck corrections can be neglected when $\ell_\mrm{imp}<2\;\upmu\mrm m$.

At higher temperature or in the presence of a non-thermal source of phonons, it may be necessary to account for the full $\mathbf{q}$-dependence in Eq.~\eqref{eq:FormFactor}.  This can be done, in principle, although the resulting temperature dependences will generally be more complicated, not described by the robust power laws we find here in the low-temperature limit.

\subsection{Direct (one-phonon) processes \label{secPhononsDirect}}

\begin{figure}
 \begin{center}
  \includegraphics{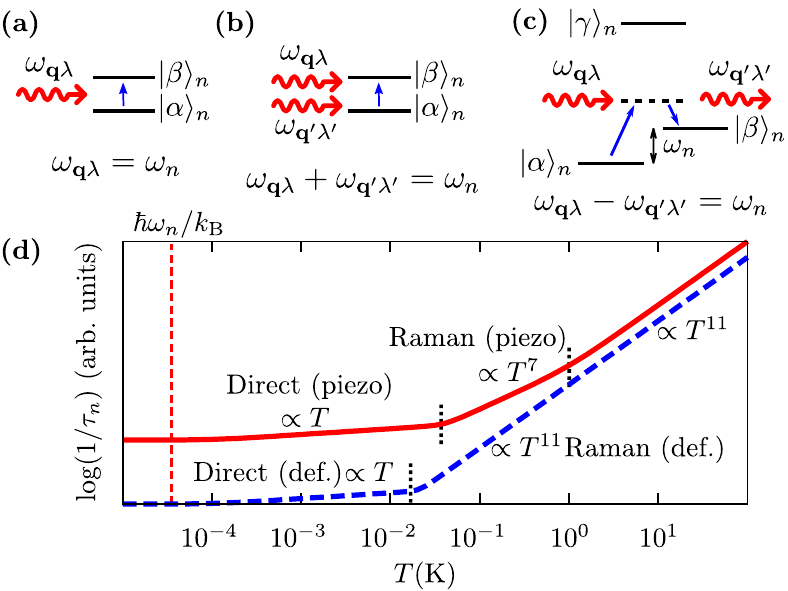}
 \end{center}
 \caption{(Color online) Coupling of a fluctuator consisting of two localized electron states interacting with a phonon bath. We consider transitions up to second order in the electron-phonon interaction. (a) Direct phonon absorption. (b) Excitation due to the two-phonon sum process. (c) Raman excitation. (a-c) We also include the corresponding relaxation processes (not shown). (d) Equilibration rate for a fluctuator coupled to phonons through the deformation and piezoelectric mechanisms, calculated with Eqs.~\eq{eqnRate1phonon} to \eq{eq:totaltaun}. Solid red line: GaAs lattice. Dashed blue line: silicon lattice. In GaAs,\cite{ioffe1998new,arlt1968piezoelectricity,yu1996fundamentals} $\Xi=a(\G_{1\mrm c})=-8.6$~eV, $e_{14}=-0.16\;\mrm{C/m^2}$, $v_\mrm{LA}=5210\;\mrm{m/s}$, $v_\mrm{TA}=3070\;\mrm{m/s}$, $\hbar\wD/\kB=360$ K, and $\ve=12.9\,\veps_0$. In silicon, $\Xi=\Xi_d+\frac13\Xi_u$, $\Xi_d=5$~eV, $\Xi_u=8.77$~eV, $e_{14}=0$, $v_\mrm{LA}=9040\;\mrm{m/s}$, $v_\mrm{TA}=5400\;\mrm{m/s}$, and $\hbar\wD/\kB=640$ K. For both GaAs and silicon, we take $|\vsq^n_{\al\bt}|=1\;\mrm{nm}$, $\sq^n_0=10\;\mrm{nm}$, $\hbar\w_n=10\;\mrm{neV}$, and $\hbar\w_\g=1\;\mrm{meV}$. \label{figPhonons}}
\end{figure}

Figure \ref{figPhonons} illustrates the fluctuator-phonon processes considered in this section. In the leading-order process, the fluctuator absorbs or emits a phonon with frequency $\w_{\mvec q\ld}=\w_n$ [see Fig.~\ref{figPhonons}(a)]. The equilibration rate corresponding to this process is obtained from the coupling Hamiltonian, Eq.~\eq{eqnElectronPhonon}, using Fermi's golden rule, Eq.~\eq{eqnFermi}. In GaAs, the conduction band has a unique minimum (a single valley), such that $\al_\chi=\dt_{\chi,1}$ in Eq.~\eq{eqntmullp}. In contrast, the conduction-band minimum of bulk silicon is six-fold degenerate.  For silicon, we take $\al_\chi=1/\sqrt{6}\;\forall\;\chi$, consistent with the ground state for donor impurities.\cite{kohn1955theory,yu1996fundamentals} Other choices of $\alpha_\chi$ would not change the final temperature dependence of the equilibration rate.  We also assume the transition dipole matrix element to be valley-independent, $\vsq_{ll'}^{\chi n}=\vsq^n_{ll'}\;\forall\;\chi$. Valley-independence of $\vsq^n_{ll'}$ amounts to neglecting anisotropy of the envelope functions $F_{\chi l}^{n}(\vecr)$ and thus of the effective mass.\cite{kohn1955theory} With the above assumptions, we find the equilibration rate for the direct process
\begin{align}
 \frac{1}{\tau^\mrm D_n}
 &=\left[\frac{1}{3}\Xi^2\left(\frac{\wn}{v_\mrm{LA}}\right)^4
     +\frac{4}{35}\left(1+\frac{4}{3}\z^2\right)
     \left(\frac{ee_{14}}{\ve}\right)^2\left(\frac{\wn}{v_\mrm{LA}}\right)^2\right]\notag\\
 &\qquad\times\frac{9\pi}{\hbar}\frac{\wn}{\mat\wD^3}|\vsqab^n|^2\coth\left(\frac{\hbar\wn}{\kB T}\right),
    \label{eqnRate1phonon}
\end{align}
where $\z=v_\mrm{LA}/v_\mrm{TA}$, with $v_\mrm{LA}$ and $v_\mrm{TA}$ the phase velocities of the longitudinal and transverse acoustic branches, respectively. Equation \eq{eqnRate1phonon} assumes the piezoelectric tensor for a zincblende-structure material, such as GaAs. For this structure, the only non-vanishing tensor element is $e_{14}$ (in Voigt notation). Silicon is not piezoelectric, resulting in $e_{14}=0$. We have also introduced the Debye frequency $\wD$, the elementary charge $e$, the mass per lattice atom $m_\mrm{at}$, and the static dielectric constant $\veps$. In GaAs, $\Xi=a(\G_{1c})\simeq-8.6$~eV, where $a(\G_{1c})$ is the volume deformation potential for the conduction-band minimum. In silicon, $\Xi=\Xi_d+\frac13\Xi_u$, where $\Xi_d$ and $\Xi_u$ are deformation potentials at zone boundaries.\cite{herring1956transport,yu1996fundamentals}

\subsection{Two-phonon processes \label{secPhononsTwo}}

We now consider the second-order processes stemming from the coupling Hamiltonian, Eq.~\eq{eqnElectronPhonon}. We first consider the two-phonon sum process. In this case, two phonons with frequencies satisfying $\wql+\w_{\mathbf{q}'\ld'}=\w_n$ are simultaneously absorbed or emitted [see Fig.~\ref{figPhonons}(b)]. We also include the Raman process, in which a phonon in mode $\mvec q\ld$ is absorbed and another is emitted in mode $\mvec q'\ld'$, with the constraint $\wql-\w_{\mvec q'\ld'}=\wn$ [see Fig.~\ref{figPhonons}(c)]. Both of these second-order processes require the presence of an auxiliary third level $\ket \g_n$, with energy splittings relative to states $\ket\al_n$ and $\ket\bt_n$ denoted by $\hbar\wag$ and $\hbar\wbg$. We obtain the effective Hamiltonians for these second-order processes using the leading-order Schrieffer-Wolff expansion, Eq.~\eq{eqnSW}, taking $\hat V^n=\hat H_\mrm{FB}^n$ from Eq.~\eq{eqnElectronPhonon}. In general, resonant denominators arise in these second-order processes for $\wql=\wag$, $\wbg$. We neglect contributions from these resonances, which are exponentially suppressed for $\hbar\wag$, $\hbar\wbg\gg\kB T$.\cite{van1940paramagnetic,yen1964phonon} We then evaluate the corresponding fluctuator equilibration rates using Fermi's golden rule, Eq.~\eq{eqnFermi}. For $\hbar\wn<\kB T$, we find the temperature and fluctuator-splitting dependences of the sum and Raman processes shown in Table~\ref{tabPhonons}. Explicitly, the equilibration rate for the Raman process is
\begin{align}
 \frac1{\tau_n^\mrm R}
  &\simeq \frac{(2\pi)^7(\sq^n_0)^4}{(\hbar\wg)^2\mat^2\wD^6}
    \left[\frac{15\pi^{4}}{11}\frac{\Xi^4}{v_\mrm{LA}^8}\left(\frac{\kB T}{\hbar}\right)^{11}\right.\notag\\
	&\qquad+ \frac{18\pi^{2}}{175}\left(\Xi\frac{ee_{14}}{\veps}\right)^2\frac{1+\frac43\z^2}{v_\mrm{LA}^6}\left(\frac{\kB T}{\hbar}\right)^{9}\notag\\
	&\qquad\left.
	+ \frac{27}{8575}\left(\frac{ee_{14}}{\veps}\right)^4\frac{(1+\frac43\z^2)^2}{v_\mrm{LA}^4}\left(\frac{\kB T}{\hbar}\right)^{7}\right],
	      \label{eqnRateRaman}
\end{align}
where we have introduced $\wg=(1/\wag+1/\wbg)^{-1}$ and $(\sq^n_0)^4=|\vsq_{\al\g}^n|^2|\vsq_{\g\bt}^n|^2+|\vsq^n_{\al\g}\cdot\vsq^{n\ast}_{\g\bt}|^2$. 

The equilibration rate for the sum process is given by Eq.~\eq{eqnRateSum} in Appendix~\ref{secSum}. Comparing Eq.~\eq{eqnRateSum} with Eq.~\eq{eqnRateRaman}, we immediately see that the prefactors are identical up to a factor of order one. Thus, using the $\w_n$ and $T$ dependences summarized in Table~\ref{tabPhonons}, the condition for the Raman process to dominate over the sum process can be shown to be $\hbar\w_n<\kBT$. In other words, the Raman process always dominates over the sum process for fluctuators that participate significantly to qubit dephasing. Thus, we neglect the sum process in the rest of this paper, regardless of the material. In contrast, the condition for the Raman process to dominate over the direct process does depend on the relevant material parameters.

\begin{table}
 \begin{center}
  \begin{tabular}{lll}
  \hline
  \hline
      & Deformation	($\sim \Xi$)		& Piezoelectric ($\sim e_{14}$)\\
   \hline
   Direct, $1/\tau^\mrm{D}_n$	& $\propto\;\wn^4\times T$		& $\propto\;\wn^2\times T$\\
   Sum, $1/\tau^\Sg_n$	& $\propto\;\wn^9\times T^2$	& $\propto\;\wn^5\times T^2$\\
   Raman	, $1/\tau^\mrm{R}_n$& $\propto\;T^{11}$		& $\propto\; T^7$\\
   \hline
   \hline
  \end{tabular}
 \end{center}
 \caption{Power-law dependences of each contribution to the fluctuator equilibration rates $1/\tau_n$ on $\wn$ and $\kBT$ for the electron-phonon interaction when $\hbar\wn<\kBT$.\label{tabPhonons}}
\end{table}

In Fig.~\ref{figPhonons}(d), we plot the total equilibration rate, 
\begin{align}\label{eq:totaltaun}
 \frac{1}{\tau_n}=\frac{1}{\tau_n^\mrm D}+\frac{1}{\tau_n^\mrm{R}},
\end{align}
as a function of temperature. The solid red (dashed blue) line shows the equilibration rate for a fluctuator in GaAs (silicon). For either material, Fig.~\ref{figPhonons}(d) illustrates a typical crossover from a low-temperature rate dominated by the direct (one-phonon) process $1/\tau_n\simeq 1/\tau_n^\mrm D \propto T$ to a high-temperature rate dominated by the two-phonon Raman process ($1/\tau_n\simeq 1/\tau_n^\mrm R \propto T^7$  for piezoelectric coupling and $1/\tau_n\simeq 1/\tau_n^\mrm R \propto T^{11}$ for deformation-potential coupling; see Table \ref{tabPhonons}).  

From Eq.~\eq{eqnRate1phonon}, the piezoelectric contribution dominates in the direct (one-phonon) process when $\wn<\w_\mrm{crit}^\mrm D$ where
\begin{align}
  \w_\mrm{crit}^\mrm D &= \sqrt{\frac{12}{35}\left(1+\frac{4}{3}\z^2\right)} \frac{e|e_{14}|}{\ve} \frac{v_\mrm{LA}}{\Xi}.
  \label{eqnwcrit}
\end{align}
For the Raman process, the piezoelectric mechanism dominates [see Eq.~\eq{eqnRateRaman}] when $T<T_\mrm{crit}$, where
\begin{align}
 T_\mrm{crit}&=\frac{1}{2\pi}\left(\frac{11}{35}\right)^{1/4}\frac{\hbar\w_\mrm{crit}^\mrm D}{\kB}.\label{eq:Tcrit}
\end{align}
From Eq.~\eqref{eq:Tcrit}, $\kB T_\mrm{crit}<\hbar\w_\mrm{crit}^\mrm D$. Thus, for fluctuators that contribute significantly to qubit dephasing (having $\hbar\omega_n<\kBT$), if the piezoelectric contribution dominates in the Raman process ($T<T_\mrm{crit}$), then it also dominates for direct absorption and emission: $\hbar\wn<\kB T<\kB T_\mrm{crit}<\hbar\w_\mrm{crit}^\mrm D$. Using the GaAs parameters given in Fig.~\ref{figPhonons}, the piezoelectric contribution then dominates in both the  direct (one-phonon) and two-phonon Raman processes if $T<1.0\;\mrm K$. Thus, in GaAs, the crossover from piezoelectric to deformation-potential mechanisms occurs at 
\begin{equation}
 T_\mrm{crit}=1.0\;\mrm{K}\qquad\mrm{[GaAs]}.	\label{eqnTcritGaAs}
\end{equation}
This feature is indeed visible in Fig.~\ref{figPhonons}(d).  Quite significantly, $T_\mrm{crit}$ depends only on material parameters and is therefore completely independent of the details of the fluctuators themselves. 

In summary, all qualitative differences between the results for GaAs and silicon in Fig.~\ref{figPhonons}(d) arise for $T<T_\mrm{crit}$(GaAs), where the piezoelectric contribution dominates in GaAs. 

%%%%%%%%%%%%%%%%%%%%%%%%%%%%%%%%%%%%%%%%%%%%%%%%%%%%%%%%%%%%%%%%%%%%%%%%%%%%%%%%%%%%%%%%%%%%%%%%%%%%%%%%

\section{Coherence time and stretching parameter from microscopic models \label{secMicro}}

In this section, we use the expressions for $\D\xi_n^2$ and $1/\tau_n$ found from microscopic models in Secs.~\ref{secElectrons} and \ref{secPhonons} to find the temperature dependences of $T_2^s$ and $\al^s$.  We first proceed numerically, which allows us to access the full temperature range. We then find explicit analytical expressions in either the slow-noise ($\tau_c\gg T_2^s$) or fast-noise ($\tau_c\lesssim T_2^s$) regime. We finally discuss implications for the interpretation of experiments.

\subsection{Numerical evaluation}

For numerical evaluation, we take the fluctuator frequency $\omega_n$ to vary inhomogeneously between fluctuators, but take all other parameters (tunnel couplings, form factors, fluctuator-qubit couplings) to be approximately independent of $n$. Taking the continuum limit of Eqs.~\eqref{eqnfs} and \eq{eqnTdM} for a large number of fluctuators [$\sum_n\to \int d\omega $, $\Delta\xi_n\to \Delta\xi(\omega)$, $\tau_n\to\tau(\omega)$] then gives
\begin{align}
 f^s(t_s)&=\frac{t_s}{\TdM}-\int_0^\infty d\w D(\w)\D\xi^2(\w)\tau^2(\w)h^s[t_s/\tau(\w)],	\label{eqnfstsNumerical}\\
 1/\TdM&=\int_0^\infty d\w D(\w)\D\xi^2(\w)\tau(\w),	\label{eqnT2MNumerical}
\end{align}
where $D(\w)$ is the fluctuator density of states. The qubit coherence time $T_2^s$ is then given directly from the numerical solution of Eq.~\eqref{eqnDefT2s} and the stretching parameter $\al^s$ is given by Eq.~\eq{eqnAlpha}. The resulting temperature dependences strongly depend on the density of states $D(\w)$. Here, we assume a near-constant density of states $D(\w)\simeq D(0)$ for $\omega \lesssim k_B T/\hbar$, where the integrand carries appreciable weight [the integral in Eq.~\eqref{eqnfstsNumerical} is cut off by $\D\xi^2(\w)\sim e^{-\hbar\omega/k_B T}$ at large frequency].\footnote{The assumption of a constant density of states in Eq.~\eqref{eqnfstsNumerical} at low temperature is consistent, e.g., with capacitance-probe spectroscopy experiments, where the inhomogeneous broadening of shallow donor levels in the dopant layer of a GaAs/AlGaAs heterostructure has been measured to be $\sim1$ meV ($1\,\mathrm{m}e\mathrm{V}/k_B\sim 10\,\mathrm{K}$).\cite{kuljanishvili2008scanning,tessmer2008nanometer} It is also a standard assumption for two-level systems in glasses.\cite{phillips1987two}}  In systems where a non-constant density of states is expected or measured, this could easily be incorporated in Eq.~\eqref{eqnfstsNumerical}, above.

\begin{figure}
 \begin{center}
  \includegraphics{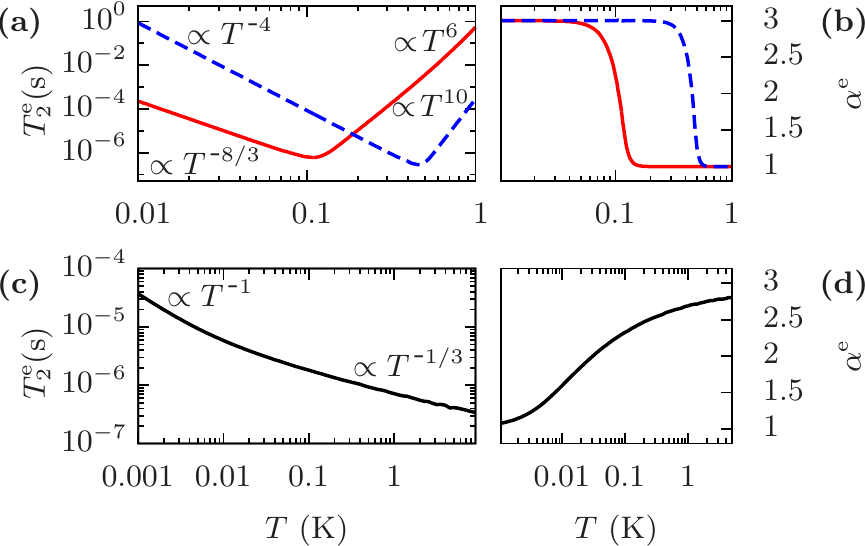}
 \end{center}
 \caption{(Color online) Hahn-echo coherence time $T_2^\mrm e$ and corresponding stretching parameter $\al^\mrm e$ (a) Coherence time from the Raman phonon process in GaAs (solid red line) and silicon (dashed blue line). (b) Corresponding stretching parameter. (c) Coherence time for the direct tunneling process. (d) Corresponding stretching parameter. (a-b) Material parameters take the same values as given in the caption of Fig.~\ref{figPhonons} for GaAs and silicon. For the Raman process, we choose $\sq_{\al\bt}^n=0$, $\sq_0^n=100\;\mrm{nm}$, $\hbar\wg=200~\upmu$eV, and $D(0)\W^2_n=1\times10^3$~$\mrm s^{-1}$ $\forall\,n$. These numbers are chosen to give $\Tde\sim1\;\upmu$s and $\al^\mrm e\simeq1$ for $T\sim100$~mK, consistent with Ref.~\onlinecite{dial2013charge}. The decay is given by a compressed exponential with $\alpha^\mathrm{e}>1$ at low temperature, but becomes exponential ($\alpha^\mathrm{e}=1$) at higher temperature. (c-d) For the direct tunneling process, we choose $D(0)\W^2_n=1\times10^2$~$\mrm s^{-1}$ and $\frac{2\pi}{\hbar}D_\mrm{el}(\e_{\al n})|t_n(\e_{\al n})|^2=10^6$~$\mrm s^{-1}$ $\forall$ $n$. The behaviors of $\Tde(\temp)$ and $\al^\mrm e(\temp)$ are radically different for the Raman mechanism and tunneling mechanism, which makes them easily distinguishable. As explained in the main text, various other qubit dephasing channels can become relevant at higher temperatures, possibly obscuring the crossovers seen here in any given experiment.\label{figDecay}}
\end{figure}

We use the numerical method described above to evaluate $T_2^\mrm e$ and $\al^\mrm e$ as a function of temperature $T$ accounting for the two-phonon Raman [Figs.~\ref{figDecay}(a),(b)] and direct-tunneling [Figs.~\ref{figDecay}(c),(d)] processes. A measurement of the distinct temperature dependences shown in Fig.~\ref{figDecay} could be used to distinguish different microscopic mechanisms.  In Figs.~\ref{figDecay}(a),(b), the red solid lines show the temperature dependences expected in GaAs, where piezoelectric coupling to phonons dominates for $T<T_\mathrm{crit}\simeq 1.0\,\mathrm{K}$, but the deformation mechanism dominates for $T>T_\mathrm{crit}$.  The blue dashed lines in Figs.~\ref{figDecay}(a),(b) show the expected behavior for silicon, where only the deformation mechanism is relevant.  The transition between distinct power-law dependences in $T_2^\mathrm{e}$ shown in Figs. \ref{figDecay}(a),(c) occurs in the crossover regime, when $\tau_c/T_2^\mathrm{e}\sim 1$.  Unlike $T_\mathrm{crit}$, discussed above, the temperature scale determining this crossover is generally non-universal, depending on the specific details of the fluctuators and their coupling to the qubit. The distinct upturn in $T_2^\mathrm{e}$ at large $T$ in Fig.~\ref{figDecay}(a) is due to motional averaging;  the Raman mechanism leads to a strong reduction in the noise correlation time at large $T$ ($\tau_c\propto 1/T^7$ or $\tau_c\propto 1/T^{11}$), which cannot be compensated by the slow growth in the noise amplitude ($\propto T$) for a constant density of states.  The result is a fast averaging of the noise and a resulting increase in coherence time $T_2^\mathrm{e}$.  It should be possible to observe such an upturn experimentally when other high-temperature qubit-dephasing mechanisms can be suppressed.  These mechanisms may arise, e.g., from direct coupling of the qubit to phonons, resulting in exponentially-activated pure dephasing from single-phonon absorption and emission,\cite{semenov2004phonon} or from strongly temperature-dependent pure-dephasing rates due to multi-phonon processes.\cite{roszak2009phonon,kornich2014phonon}

For all processes investigated here, there is a crossover, as a function of temperature, from the fast-noise (Markovian) limit, $\tau_c/T_2^\mathrm{e}\ll 1$, in which $\alpha^\mathrm{e}\simeq 1$, to the slow-noise limit, $\tau_c/T_2^\mathrm{e}\gg 1$, where $\alpha^\mathrm{e}\simeq 3$ (see Sec.~\ref{secCompressed}).  Strikingly, for the Raman process, the crossover is from the slow-noise to the fast-noise limit with increasing temperature [Fig.~\ref{figDecay}(b)].  In contrast, the tunneling process leads to a crossover from fast- to slow-noise with increasing temperature [Fig.~\ref{figDecay}(d)].  In the case of the Raman process, the fast-noise limit is naturally reached at large temperature because of the rapid decrease of the noise correlation time ($\tau_c\sim 1/T^7$ or $\tau_c\sim 1/T^{11}$) in combination with an increase in $T_2^s$ due to motional averaging (see the discussion above).  For the tunneling process, the correlation time saturates at high temperature $\tau_c\sim \tau_n\propto \mathrm{const.}$ (see Table \ref{tabTunnel}), while the amplitude of the noise increases as progressively more fluctuators satisfying $\hbar\omega_n\lesssim k_B T$ contribute, leading to a decrease in $T_2^s$ and a corresponding transition to the slow-noise limit $\tau_c/T_2^s\gg 1$ at high temperature.

\begin{table*}
 \begin{center}
   \begin{tabular}{l|c|ll|c|l}
    \hline\hline
      & Fast noise (Markovian, $\tau_c\ll T_2^s$)		& \multicolumn{2}{c|}{Slow noise ($\tau_c\gg T_2^s$)}	& Crossover ($\tau_c\lesssim T_2^s$)	& \\
      Process	& $1/T_2^\ast=1/\Tde=1/\TdM$ & $1/T_2^\ast$	& $1/T_2^\mrm e$	& $\bt^s\propto\eta$	& $\al^s$\\
    \hline
      Direct tunneling	& $\propto T$	& $\propto T^{1/2}$	& $\propto T^{1/3}$	& $\propto T$	& $\upa$\\
      Cotunneling		& $\propto1$	& $\propto T^{1/2}$	& $\propto T^{2/3}$	& $\propto T^{-1}$	& $\dwna$\\
    \hline
      Direct (deformation)
      & $\propto T^{-17/2}$	& $\propto T^{1/2}$	& $\propto T^2$		& $\propto T^{-39/2}$	& $\dwna$\\
      Direct (piezoelectric)
      & $\propto T^{-4}$	& $\propto T^{1/2}$	& $\propto T^{4/3}$	& $\propto T^{-11}$ 	& $\dwna$\\
      Raman (deformation)	& $\propto T^{-10}$	& $\propto T^{1/2}$	& $\propto T^{4}$	& $\propto T^{-21}$ & $\dwna$\\
      Raman (piezoelectric)	& $\propto T^{-6}$	& $\propto T^{1/2}$	& $\propto T^{8/3}$	& $\propto T^{-13}$ & $\dwna$\\
    \hline\hline
   \end{tabular}
   \vspace{5mm}
 \end{center}
 \caption{Temperature dependence of the coherence factor $C^s(t_s)\simeq\exp[-(t_s/T_2^s)^{\al^s}]$ for a qubit coupled to fluctuators interacting with either an electron bath (first two rows) or a phonon bath (last four rows). We give the coherence-time temperature dependences for both free-induction decay ($s\rightarrow\ast$) and Hahn echo ($s\rightarrow\mrm e$) in the limits of fast noise (Markovian, $\tau_c\ll T_2^s$) and slow noise ($\tau_c\gg T_2^s$). In the crossover regime, we also give the temperature dependence of $\bt^s=\al^s-1$ for $\tau_c\lesssim T_2^s$. The last column indicates whether $\al^s$ increases ($\upa$) or decreases ($\dwna$) as a function of temperature $T$. All these results are obtained for a near-constant fluctuator density of states $D(\w)\simeq D(0)$ for $\w\lesssim\kB T/\hbar$. Different densities of states could easily be accounted for using Eqs.~\eq{eq:T2Star} to \eq{eqnPropbeta}. Predictions for the two-phonon sum process are absent since, for $\hbar\wn<\kBT$, these processes are always negligible relative to the Raman processes (see Sec.~\ref{secPhonons}). \label{tabT2}}
\end{table*}

\subsection{Slow- and fast-noise regimes}

As described above, given sufficient microscopic information, it is possible to make quantitative predictions for the temperature dependence of the qubit coherence time $T_2^s$ and stretching parameter $\alpha^s$.  To do this, we would need a good description of the relevant transition dipole matrix elements $\vsq^n_{\alpha\beta}$ or tunnel couplings $t_{\alpha n}(\epsilon)$ as well as the fluctuator density of states and microscopic material-specific parameters.  When the specific impurities associated with charge noise can be identified, it may be possible to estimate or measure these quantities.  In many experiments, however, it may be difficult to establish the specific source of charge noise and the associated parameters.  In this case, we can still make strong analytical predictions about the scaling of $T_2^s$ with temperature in either the fast-noise ($\tau_c\lesssim T_2^s$) or slow-noise ($\tau_c \gg T_2^s$) regime.  

We allow the qubit-fluctuator couplings $\W_n$, dipole matrix elements $\vsq^n_{\al\bt}$, etc. to vary generally with $n$. However, to make analytical progress, we assume that these parameters are approximately independent of $\w_n$ for $\w_n\lesssim\kBT/\hbar$ where $\Delta\xi^2(\omega_n)$ is appreciable.  To determine the simple scaling behavior, we replace the exponential dependence $\Delta\xi^2(\omega_n)\sim e^{-\hbar\omega_n/\kBT}$ with a hard cutoff at $\hbar\omega_n = \kBT$. Taking the continuum limit of Eq.~\eqref{eqnTdM} for the fast-noise limit $(\tau_c\ll T_2^s)$ then gives
\begin{equation}
 \frac1\Tdst=\frac1\Tde=\frac{1}{\TdM}\propto\int_0^{\kBT/\hbar}d\w\;D(\w)\tau(\w,T).\label{eqnPropM}
\end{equation}
With the same assumptions, we perform the continuum limit in Eqs.~\eqref{eqnTdNMst} and \eqref{eqnTdNMe} for the slow-noise limit $(\tau_c\gg T_2^s)$, giving 
\begin{eqnarray}
 \frac{1}{\Tdst}&\propto &\left[\int_0^{\kB T/\hbar} d\w\;D(\w)\right]^{1/2},\label{eq:T2Star}\\
 \frac{1}{\Tde}&\propto &\left[\int_0^{\kBT/\hbar}d\w\frac{D(\w)}{\tau(\w,T)}\right]^{1/3}.	\label{eqnPropNM}
\end{eqnarray}
From Eqs.~\eq{eqnBetaStar} to \eq{eqneta} for $\bt^s$ and $\eta$, we also have, in the fast-noise regime
\begin{align}
 \bt^s\propto\eta\propto\int_0^{\kB T/\hbar}d\w\;D(\w)\tau^2(\w,T).	\hspace{7mm}(\tau_c\lesssim T_2^s)	\label{eqnPropbeta}
\end{align}

In the slow-noise limit, the inhomogeneously broadened decay time $T_2^*$ is independent of the fluctuator equilibration time $\tau_n$. This decay time is therefore independent of the specific microscopic mechanism giving rise to fluctuator dynamics and can be used to measure the frequency dependence of the fluctuator density of states. Indeed, taking $D(\w)=D_0\w^a$, Eq.~\eq{eq:T2Star} gives
\begin{align}
 1/\Tdst\propto T^{\frac{a+1}{2}},	\hspace{2cm}[\tau_c\gg\Tds]	\label{eqnT2sDOS}
\end{align}
where we have assumed $a>-1$. Thus, the scaling with temperature of $1/\Tdst$ in the slow-noise regime can be used to determine $a$ under the assumption that fluctuator parameters other than $\w$ (i.e. $\Omega_n$, $\vsq_{\al\bt}^n$, etc.) are approximately frequency-independent for $\w\lesssim\kB T/\hbar$.

In Tables~\ref{tabTunnel} and \ref{tabPhonons}, we give the $\w$ and $T$ dependences of $1/\tau$ for all fluctuator-bath processes considered in this paper. Substituting these dependences into Eqs.~\eq{eqnPropM} to \eq{eqnPropNM} and assuming a constant fluctuator density of states ($a=0$) gives the power-law scalings for $T_2^s$ shown in Table~\ref{tabT2}. These scalings are consistent with those obtained numerically in Fig.~\ref{figDecay}. Similar tables could easily be built for different values of $a$, i.e., for non-constant fluctuator densities of states. 

\begin{figure}
 \begin{center}
    \includegraphics[width=0.47\textwidth]{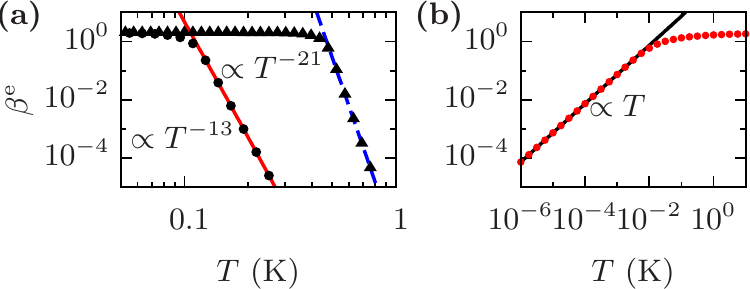}
 \end{center}
 \caption{(Color online) Parameter $\bt^\mrm e=\al^\mrm e-1$. (a) Raman process. Black circles (black triangles): $\bt^\mrm e$ for GaAs (silicon), from the numerical method of Eqs.~\eq{eqnDefT2s}, \eq{eqnAlpha}, and Eqs.~\eq{eqnfstsNumerical}, \eq{eqnT2MNumerical}. Solid red line (dashed blue line): analytical temperature dependence for $\tau_c\lesssim T_2^s$ from Table~\ref{tabT2} for GaAs (silicon). (b) Direct tunneling. Red circles: numerical method. Solid black line: analytical prediction. Assumptions and microscopic parameters are the same as in Fig.~\ref{figDecay}. \label{figBeta}}
\end{figure}

In Fig.~\ref{figBeta}, we plot $\bt^\mrm e=\al^\mrm e-1$ as a function of temperature for the Raman process [Fig.~\ref{figBeta}(a)] and direct tunneling [Fig.~\ref{figBeta}(b)]. We evaluate Eqs.~\eq{eqnfstsNumerical} and \eq{eqnT2MNumerical} numerically with the same assumptions and parameters as described in the caption of Fig.~\ref{figDecay}. These numerical results are represented in Fig.~\ref{figBeta} by circles and triangles. The analytical predictions of Table~\ref{tabT2} are also plotted as straight lines. As expected from the discussion above Eq.~\eq{eqnT2StarFast}, these analytical results only substantially deviate from exact numerical calculations when $\bt^\mrm e\simeq3\eta\sim1$, corresponding to $\tau_c\sim\Tde$. Indeed, when $\eta\rightarrow\infty$ (the slow-noise limit), Eq.~\eq{eqnPropbeta} predicts an unbounded growth of $\bt^s$, while, from Eqs.~\eq{eqnTdNMst} and \eq{eqnTdNMe}, $\bt^s$ saturates to $1(2)$ for free-induction decay (Hahn echo). However, for $\tau_c> T_2^s$, Eq.~\eq{eqnPropbeta} and the corresponding power laws in Table~\ref{tabT2} still give the trends in $\al^s$ [increasing ($\upa$) or decreasing ($\dwna$)] shown in Table~\ref{tabT2}. 

In Table~\ref{tabT2}, all processes we have considered can be distinguished from a combined measurement of the temperature dependence of $T_2^\mrm s$ and $\bt^s$. From this table, it should be possible to rule out specific fluctuator noise mechanisms based on a measurement of $T_2^s$ and $\al^s$ as a function of temperature.

\subsection{Relevance to experiment \label{secExp}}

To assess the usefulness of the approach described here, we now consider an application to a recent experiment.  In Ref.~\onlinecite{dial2013charge}, Dial \emph{et al.} have observed coherence decay as a function of temperature for a qubit defined by singlet and triplet spin states in a two-electron double quantum dot in GaAs.  These measurements revealed an approximate linear dependence of the inhomogeneously broadened decay time, $\Tdst\propto A-B\temp$, with $\al^\ast=2$ for temperatures between $\sim 50$~mK and $\sim250$~mK. This behavior may be compatible with any dependence $\Tdst\propto1/T^{\frac{a+1}{2}}$ given by Eq.~\eqref{eqnT2sDOS} with $a>-1$. Thus, a more precise measurement of $T_2^\ast$ as a function of temperature may establish the specific form of the fluctuator density of states in this experiment.  

Under the assumption of a constant fluctuator density of states [$D(\w)\sim\w^a$ with $a=0$], we attempt to apply the results of Table~\ref{tabT2} to describe the experimental results of Ref.~\onlinecite{dial2013charge}. In Ref.~\onlinecite{dial2013charge}, the authors measured $T_2^\mrm e(T)$ and $\al^\mrm e(T)$ and found that: (i) $T_2^\mrm e(T)\propto T^{-\g}$ with $\g\sim2$ for the whole temperature range of the experiment, (ii) $\bt^\mrm e$ decreases monotonically as $T$ increases from $\sim$ 50 mK to $\sim$ 150 mK, and (iii) $\bt^\mrm e\lesssim0.7$ for the whole temperature range, corresponding to the fast-noise regime, in which $\tau_c\lesssim T_2^\mrm e$. In this regime, Eqs.~\eq{eqnT2eFast} and \eq{eqnBetaE} yield $T_2^\mrm e\simeq T_\mrm{2M}/(1+\bt^\mrm e)\sim T_\mrm{2M}$ up to a correction $\mathcal O(\bt^\mrm e\TdM)$. The first column of Table~\ref{tabT2} should then accurately reflect the trend in $\Tde(T)$ in the fast-noise regime, consistent with $\beta^\mrm e\to0$. For all phonon mechanisms, we find that $T_2^\mrm e$ increases with temperature for $\tau_c\lesssim T_2^\mrm e$, while the data from Ref.~\onlinecite{dial2013charge} exhibit the opposite trend. The only mechanism in Table~\ref{tabT2} for which $T_2^\mrm e$ correctly decreases when $T$ increases in the fast-noise regime is direct tunneling. However, for this process, $\bt^\mrm e$ increases monotonically with temperature, in contradiction with the experimental data of Ref.~\onlinecite{dial2013charge}. Therefore, under the assumption of a constant density of states, none of the physical processes displayed in Table~\ref{tabT2} can, alone, explain all the observations listed above.

One of the assumptions behind Table~\ref{tabT2} may be violated in the context of Ref.~\onlinecite{dial2013charge}. Here, we review the assumptions and limitations leading to this table. To begin with, it may be that the true fluctuator density of states was not constant in the experiment of Ref.~\onlinecite{dial2013charge}. A precise measurement of $T_2^\ast$ in the slow-noise regime can be used to establish the true frequency dependence of the fluctuator density of states through Eq.~\eq{eqnT2sDOS}. In addition, for phonon mechanisms, we have assumed a long-wavelength limit to establish the low-frequency behavior of the fluctuator equilibration rates. From Eq.~\eq{eqnab}, this assumption may be violated for fluctuators with large extended orbital states, or at high temperatures, leading to phonon-bottleneck effects.\cite{benisty1995reduced} 
Finally, we have assumed that the dominant dephasing mechanism results from coupling to charge fluctuators.  It is, of course, possible that other decay channels become relevant. For example, in the presence of an independent extrinsic Markovian dephasing process, the coherence factor takes the form
\begin{align}
 C^\mrm e(t_s)=\exp\left[-\frac{t_s}{T'_2}-\left(\frac{t_s}{T_2^s}\right)^{\alpha^s}\right].\label{eqnMulti}
\end{align}
In the above equation, $T_2^s$ and $\al^s$ are the decay time and stretching parameter for the fluctuator processes presented here, while $T'_2$ is the decay time due to an additional Markovian dephasing process acting directly on the qubit. At high temperature, many extrinsic dephasing mechanisms (not related to charge fluctuators) may become relevant (these may be due, e.g., to coupling to phonons\cite{mccumber1963linewidth,roszak2009phonon,kornich2014phonon}). The first term in Eq.~\eq{eqnMulti} may then dominate over the second. To ensure that the fluctuator mechanisms presented in this paper are the dominant source of dephasing, it may be necessary to understand and suppress alternative sources of dephasing (by, e.g., working at sufficiently low temperature).  Alternatively, when these alternate sources of dephasing are well understood, a combined formula such as Eq.~\eq{eqnMulti} could be used to extract the values of $T_2^s$ and $\alpha^s$ associated with fluctuator dynamics, even in the presence of extrinsic dephasing mechanisms.

To further illustrate how Eq.~\eq{eqnMulti} can be used to identify interactions at the origin of fluctuator dynamics, we apply it to the analysis of the data from Ref.~\onlinecite{dial2013charge}. We take $T_2^\mrm e$ to be the Hahn-echo decay time for one of the fluctuator processes of Table~\ref{tabT2} in the slow-noise limit (in which $\al^\mrm e=3$). When $T'_2< T_2^\mrm e$, the contribution to qubit decay of the extrinsic Markovian process dominates over the contribution of the fluctuators. We then find the qubit decay time $T_2$ including both fluctuator and extrinsic processes. We do so by setting the argument of the exponential in Eq.~\eq{eqnMulti} equal to one and solving for $t_s\equiv T_2$ using an expansion in increasing powers of $T'_2/T_2^\mrm e$. Substituting the resulting expression for $T_2$ in the definition of the stretching parameter $\al$, Eq.~\eq{eqnAlpha}, we find the form of $\bt$ including both processes (fluctuator and extrinsic) to leading order in $T'_2/T_2^\mrm e$. We take $T_2\simeq T'_2\propto T^{-\dt}$ for the extrinsic dephasing mechanism, with $\dt$ the exponent obtained from the experiment of Ref.~\onlinecite{dial2013charge}, and $T_2^\mrm e\propto T^{-\g}$, with $\g$ the appropriate exponent for the relevant fluctuator mechanism from Table~\ref{tabT2}. We then find, to leading order in $T'_2/T_2^\mrm e$,
\begin{align}
 \bt\propto T^{3(\g-\dt)}	.	\label{eqnGamma}
\end{align}
The decreasing trend for $\bt(T)$ observed in Ref.~\onlinecite{dial2013charge} from $\sim50$~mK to $\sim150$~mK is thus reproduced for $\g<\dt$. For $\dt=2$, as written in Ref.~\onlinecite{dial2013charge}, the decreasing trend for $\bt(T)$ is consistent with all the fluctuator mechanisms from Table~\ref{tabT2} in the slow-noise limit except for the Raman processes (from either piezoelectric or deformation mechanisms). However, for $100$~mK$<T<200$~mK, Kornich \etal~have predicted Markovian decay of singlet-triplet coherence at a rate $\propto T^{3}$ due to two-phonon processes including spin-orbit coupling (see Fig.~3 of Ref.~\onlinecite{kornich2014phonon}). This behavior is compatible with the experimental data of Ref.~\onlinecite{dial2013charge} in the relevant temperature range ($100$~mK$<T<200$~mK). Taking $\dt=3$ in Eq.~\eq{eqnGamma} implies that the observed decreasing trend for $\bt(T)$ with $T<100$~mK becomes compatible with all the fluctuator mechanisms in Table~\ref{tabT2} except the Raman process due to deformation coupling to phonons. With the help of Eq.~\eq{eqnGamma} and knowing $\dt$ from a precise measurement of $T_2(T)$ in the fast-noise regime, $\g$ could be estimated through a precise measurement of $\bt$ as a function of $T$, allowing for further identification of fluctuator processes.

%%%%%%%%%%%%%%%%%%%%%%%%%%%%%%%%%%%%%%%%%%%%%%%%%%%%%%%%%%%%%%%%%%%%%%%%%%%%%%%%%%%%%%%%%%%%%%%%%%%%%%%%

\section{Conclusions}\label{secConclusions}

We have described qubit dephasing due to two-level fluctuators undergoing equilibration dynamics with either electron or phonon reservoirs.  Even for a Lorentzian noise spectrum, which arises naturally for two-level fluctuators, the qubit coherence factor is well approximated by a compressed exponential $\exp[-(t_s/T_2^s)^{\al^s}]$. In contrast with the situation for $1/f$ noise,\cite{cywinski2008how,medford2012scaling} here the stretching parameter $\al^s$ depends on the chosen pulse sequence $s$ and obeys a universal relation, $(\al^\mrm e-1)/(\al^\ast-1)\simeq3$, in the fast-noise regime, in which $T_2^s\gtrsim\tau_c$.

We have determined the explicit temperature dependences for the stretching parameter $\al^s$ and coherence time $T_2^s$ from several microscopic mechanisms giving rise to fluctuator equilibration dynamics.  These mechanisms include direct tunneling and cotunneling between localized electronic states and an electron reservoir. We have also considered coupling of two-level charge fluctuators to a phonon bath. In the latter case, we have allowed for direct phonon absorption and emission, as well as the two-phonon sum and Raman processes. We have found that different fluctuator-bath processes lead to distinct temperature dependences for $T_2^s$ and $\al^s$. A measurement of the predicted temperature dependences should thus allow to experimentally distinguish between physical processes at the origin of fluctuator noise, providing an additional tool to suppress charge noise.

\begin{acknowledgments}
  We acknowledge financial support from NSERC, CIFAR, FRQNT, INTRIQ, and the W.~C.~Sumner Foundation.
\end{acknowledgments}

%%%%%%%%%%%%%%%%%%%%%%%%%%%%%%%%%%%%%%%%%%%%%%%%%%%%%%%%%%%%%%%%%%%%%%%%%%%%%%%%%%%%%%%%%%%%%%%%%%%%%%%%
%%%%%%%%%%%%%%%%%%%%%%%%%%%%%%%%%%%%%%%%%%%%%%%%%%%%%%%%%%%%%%%%%%%%%%%%%%%%%%%%%%%%%%%%%%%%%%%%%%%%%%%%
%%%%%%%%%%%%%%%%%%%%%%%%%%%%%%%%%%%%%%%%%%%%%%%%%%%%%%%%%%%%%%%%%%%%%%%%%%%%%%%%%%%%%%%%%%%%%%%%%%%%%%%%
\appendix

%%%%%%%%%%%%%%%%%%%%%%%%%%%%%%%%%%%%%%%%%%%%%%%%%%%%%%%%%%%%%%%%%%%%%%%%%%%%%%%%%%%%%%%%%%%%%%%%%%%%%%%%

\section{Corrections to the leading-order Magnus expansion and the Gaussian approximation \label{secMagnus}}

The discussion in Sec.~\ref{secTwoLevel} applies when the fluctuator dynamics are well described by the leading term in the Magnus expansion under a Gaussian approximation. In this Appendix, we derive the leading-order corrections to the formulas of Sec.~\ref{secTwoLevel} and find a simple condition for which these corrections can safely be neglected.

Using the Magnus expansion, the interaction-picture time-evolution operator corresponding to the perturbation given by Eq.~\eq{eqnHQF} is (taking $t\equiv t_s$ to simplify the notation)
\begin{equation}
 \hat U(t)=\exp\left[-i\sum_{m=1}^\infty\hMm(t)\right].
\end{equation}
Here, the term $\hMm(t)$ results from $m$ integrals of $m-1$ nested commutators involving the interaction-picture perturbation $V'_\mrm I(t)$.  When the perturbation is sufficiently weak, contributions with large $m$ will be suppressed since $\hMm(t)=\mathcal{O}\left(\hat V'_\mrm I(t)^m\right)$. Explicit expressions for the first few orders of the Magnus expansion can be found in the literature.\cite{maricq1982application,burum1981magnus,blanes2009magnus} To evaluate the coherence dynamics of the qubit, we calculate 
\begin{equation}
 \mean{\spl(t)}=\eul{i\phi(t)}\tr[\hat U^\dagger(t)\spl\hat U(t)\hat\rho(0)],\label{eqnSigmaPlus}
\end{equation}
with $\phi(t)$ given by 
\begin{align}
 \phi(t)&=\wQ'\int_0^{t}dt_1s(t_1),	\label{eqnphi}\\
 \wQ'&=\wQ+\sum_n\W_n\mean{\tz_n}.
\end{align}
Even-order terms in the Magnus expansion are proportional to $(\sz)^{2m}=1$ while odd-order terms are proportional to $(\sz)^{2m+1}=\sz$. Thus, we have
\begin{align}
 &\hat U^\dagger(t)\spl\hat U(t)=\eul{i\liouv_\mrm M(t)}\spl,\\
 &\liouv_\mrm M(t)\cdot=\frac 12\commut{\hxodd(t)\sz+\hxeven(t)}{\cdot},
\end{align}
where we have introduced $\hxodd(t)=\sum_{m=0}^\infty \ham_\xi^{(2m+1)}(t)$ and $\hxeven(t)=\sum_{m=0}^\infty \ham_\xi^{(2m)}(t)$. Each $\ham_\xi^{(m)}(t)$ is the $m$-th order term in the Magnus expansion associated with $\hat{V}_\xi(t) \equiv s(t)\hbar\hxi(t)$. Expanding $\eul{i\liouv_\mrm M(t)}$ in a Taylor series around $\liouv_\mrm M(t)=0$ and applying every resulting term on $\spl$, we find a recursion relation that leads to
\begin{align}
 &\hat U^\dagger(t)\spl\hat U(t)=\eul{i\liouv_\xi(t)}\spl,\label{eqnUdsigmaU}\\
 &\liouv_\xi(t)\cdot=\frac 12\left[\left\{\hxodd(t),\cdot\right\}+\commut{\hxeven(t)}{\cdot}\right].\label{eqnLxi}
\end{align}
Crucially, $\liouv_\xi(t)$ does not act on the space of qubit operators.  We define the coherence factor $\tilde{C}(t)$ through 
\begin{equation}
 \mean{\spl(t)}=\mean{\spl(0)}\tilde{C}(t),
\end{equation}
where $\tilde{C}(t)$ contains phase information, and is related to the coherence factor given in the main text through $C(t) = |\tilde{C}(t)|$.  For an initially separable state $\hat\rho(0)=\hat\rho_\mrm Q(0)\otimes\hat\rho_\mrm{FB}(0)$, Eq.~\eq{eqnSigmaPlus} combined with Eq.~\eq{eqnUdsigmaU} then gives
\begin{align}
 &\tilde{C}(t)=\eul{i\phi(t)}\mean{\eul{i\liouv_\xi(t)}},\\
 &\mean{\eul{i\liouv_\xi(t)}}\equiv\tr_\mrm{FB}\left[\left(\eul{i\liouv_\xi(t)}1\right)\hat\rho_\mrm{FB}(0)\right].
\end{align}
Since the qubit experiences noise due to many uncorrelated fluctuators, we expect a cumulant expansion to converge rapidly.  To perform the  cumulant expansion, we rewrite $\mean{\eul{i\liouv_\xi(t)}}$ in terms of $\delta \liouv_\xi(t) = \liouv_\xi(t)-\mean{\liouv_\xi(t)}$:
\begin{eqnarray}
 \mean{\eul{i\liouv_\xi(t)}}=\eul{i\mean{\liouv_\xi(t)}}\mean{\eul{i\delta\liouv_\xi(t)}}
 	=\eul{i\mean{\liouv_\xi(t)}} e^\chi, \\
 \chi=\sum_{p=1}^\infty \frac{(-1)^{p+1}}{p}\!\!\!\sum_{m_1,...,m_p}\prod_{i=1}^p\frac{\bmean{[i\delta\liouv_\xi(t)]^{m_i}}}{m_i!},\label{eqnCumulant}
\end{eqnarray}
where the sums over $m_i$ range from 1 to $\infty$. Eq.~\eq{eqnCumulant} defines an expansion in increasing powers of $\liouv_\xi(t)$, while $\liouv_\xi(t)$ is itself obtained from the Magnus expansion associated with $s(t)\hbar\hat\xi(t)$. Terms of common powers of $\hat\xi(t)$ can then be collected. Up to and including $\mathcal O(\hat\xi(t)^4)$, we find
\begin{equation}
 \tilde{C}(t)=\eul{i[\phi(t)+\phi_3(t)]}\exp\left[-f_2(t)-f_4(t)\right],
\end{equation}
where $\phi(t)$ is given by Eq.~\eq{eqnphi},
\begin{align}
 \phi_3(t)&=\bmean{\hamxi{3}(t)}+\sfrac i4\bmean{\left[\hamxi{2}(t),\hamxi{1}(t)\right]}\notag\\
	  &\hspace{35mm}-\sfrac16\bmean{[\hamxi{1}(t)]^3},
\end{align}
and
\begin{align}
 f_2(t)&=\sfrac12\bmean{[\hamxi{1}(t)]^2},	\label{eqnLeading}\\
 f_4(t)&=	-\sfrac18\left[\sfrac13\bmean{[\hamxi{1}(t)]^4}-\bmean{[\hamxi{1}(t)]^2}^2\right]	\notag\\
  &\qquad+\sfrac12\bmean{\left\{\hamxi{3}(t),\hamxi{1}(t)\right\}}			\notag\\
  &\qquad+\sfrac{i}{24}\bmean{\left\{\hamxi{1}(t),\left[\hamxi{2}(t),\hamxi{1}(t)\right]\right\}}		\notag\\
  &\qquad+\sfrac{i}{12}\bmean{\left[\hamxi{2}(t),[\hamxi{1}(t)]^2\right]}.	\label{eqnSubleading}
\end{align}
The leading-order term in the expansion of Eq.~\eq{eqnCumulant} is given by Eq.~\eq{eqnLeading} and corresponds to the first-order Magnus expansion under the Gaussian approximation. Eq.~\eq{eqnSubleading} gives the first subleading term in $|\tilde C(t)|$. The corrections given by Eq.~\eq{eqnSubleading} come both from the non-Gaussian nature of $\hat\xi(t)$ and from the fact that $\hat\xi(t)$ does not generally commute with itself at different times. 

All the terms in Eq.~\eq{eqnSubleading} involve correlators of the form $\mean{\hat\xi_{n_1}(t_1)\hat\xi_{n_2}(t_2)\hat\xi_{n_3}(t_3)\hat\xi_{n_4}(t_4)}$. Following from the definition given in Eq.~\eq{eqnXi} when the initial state of the fluctuators is factorizable [$\rho_\mrm{FB}(0)=\prod_n\rho^n_\mrm{FB}(0)$], the operators $\hat\xi_n(t)$ for the noise due to single fluctuators have the following properties 
\begin{align}
 &\mean{\hxi_n(t)}=0,	\label{eqnPropxi1}\\
 &\mean{\hxi_{n_1}(t_1)\hxi_{n_2}(t_2)}=\dt_{n_1,n_2}\mean{\hxi_{n_1}(t_1)\hxi_{n_1}(t_2)},	\label{eqnPropxi2}\\
 &\mean{[\hxi_{n}(t_1),\hxi_{n}(t_2)]}=0.	\label{eqnPropxi3}
\end{align}
The last property in Eq.~\eq{eqnPropxi3} comes from Eq.~\eq{eqnOU}. In addition, when the evolution of each fluctuator is given by a Markovian master equation of the form of Eq.~\eq{eqnLiouv}, we find from the standard formula for multitime averaging\cite{gardiner2000quantum}
\begin{align}
 &\mean{\hat\xi_n(t_1)\hat\xi_n(t_2)\hat\xi_n(t_3)\hat\xi_n(t_4)}\notag\\
 &=\left[\D\xi_n^4+\D\xi'^4_n\eul{-|t_2-t_3|/\tau_n}\right]\eul{-|t_1-t_2|/\tau_n}\eul{-|t_3-t_4|/\tau_n}\label{eqng4exact}
\end{align}
where we have introduced
\begin{align}
 \D\xi'^4_n=16\W_n^4\frac{\gup\gdwn(\gup-\gdwn)^2}{(\gup+\gdwn)^4}.\label{eqnDXiprime}
\end{align}
Substituting the detailed-balance relation $\gup/\gdwn=\exp(-\hbar\wn/\kBT)$ into Eq.~\eq{eqnDXiprime}, which we substitute again in Eq.~\eq{eqng4exact}, we find an approximate upper bound for the fourth-order correlation function in Eq.~\eq{eqng4exact}:
\begin{align}
 &\mean{\hat\xi_n(t_1)\hat\xi_n(t_2)\hat\xi_n(t_3)\hat\xi_n(t_4)}\lesssim\D\xi^4_n\eul{-|t_1-t_2|/\tau_n}\eul{-|t_3-t_4|/\tau_n},	\label{eqng4}
\end{align}
for $\hbar\wn<\kBT$, neglecting factors of order 1. In Eq.~\eq{eqng4}, $\D\xi_n$ is given by Eq.~\eq{eqnsech}. Substituting Eqs.~\eq{eqnPropxi1} to \eq{eqnPropxi3} and Eq.~\eq{eqng4} into Eq.~\eq{eqnSubleading}, we find expressions for the upper bound on $f_4^s(t)$, the first subleading correction to $|\tilde C(t)|$ for dynamical-decoupling sequence $s$. Taking the fast-noise limit ($\tau_c\ll T_2^s$), and taking a typical value $t\sim T_2^s$, we drop exponentially small corrections in $t/\tau_n\sim T_2^s/\tau_c$. The inequality for $f_4^s(t)$ then becomes, for both free-induction decay and Hahn echo,
\begin{align}
 f_4^s(t)\lesssim\sum_n\D\xi_n^4\tau_n^3t \;\forall\;s.\hspace{2.1cm}(\tau_c\ll T_2^s)	\label{eqnbndf4fast}
\end{align}
In the opposite, slow-noise limit ($\tau_c\gg T_2^s$), we expand the upper bound on $f_4^s(t)$ in a Taylor series around $t/\tau_n=0$. Keeping only the leading term, we find
\begin{align}
 f_4^\ast(t)&\lesssim\sum_n\D\xi_n^4t^4,	\hspace{2.3cm}(\tau_c\gg T_2^s)	\label{eqnbndf4slowfid}\\
 |f_4^\mrm e(t)|&\lesssim\sum_n\frac{\D\xi_n^4t^5}{\tau_n}.			\label{eqnbndf4slowecho}
\end{align}
Typically, approximately $N$ fluctuators will contribute to qubit dephasing, with $N$ defined by
\begin{equation}
 N\equiv\frac{\left(\sum_n\D\xi_n^2\right)^2}{\sum_n\D\xi_n^4}=4\frac{(1/T_\mrm{2,sl.}^\ast)^4}{\sum_n\D\xi_n^4},	\label{eqnN}
\end{equation}
where $T_\mrm{2,sl.}^\ast$ is the free-induction decay time in the slow-noise limit, Eq.~\eq{eqnTdNMst}. Assuming that $\tau_n$ varies slowly with $n$ for $\hbar\wn\lesssim\kBT$, we replace $\tau_n\rightarrow\tau_c$ in Eqs.~\eq{eqnbndf4fast} to Eq.~\eq{eqnbndf4slowecho}, $\tau_c$ being given by Eq.~\eq{eqnTauc}. Also using the definition of $N$ given by Eq.~\eq{eqnN}, we find
\begin{align}
 f_4^s(t)&\lesssim\frac{1}{N}\left(\frac{\tau_c}{T_\mrm{2,sl.}^\ast}\right)^3\frac{t}{T_\mrm{2,sl.}^\ast}\;\forall\;s,
	  \hspace{1cm}(\tau_c\ll T_2^s)	\label{eqnf4fast}\\
 f_4^\ast(t)&\lesssim\frac{1}{N}\left(\frac{t}{T_\mrm{2,sl.}^\ast}\right)^4,
	  \hspace{2.35cm}(\tau_c\gg T_2^s)	\label{eqnf4stslow}\\
 f_4^\mrm e(t)&\lesssim\frac{1}{N}\frac {t^5}{(T_\mrm{2,sl.}^\ast)^4\tau_c}.
						\label{eqnf4eslow}
\end{align}
As explained in Sec.~\ref{secCompressed}, the leading term $f_2(t)$ in the combined Magnus and cumulant expansion is well approximated by $f_2(t)\simeq(t/T_2^s)^{\al_s}$. This leading term then dominates over the subleading contribution given by Eqs.~\eq{eqnf4fast} to \eq{eqnf4eslow} when (taking $t\sim\Tds$ and neglecting factors of order 1)
\begin{align}
 &N\gg1\;\forall\;s,	\hspace{41mm}(\tau_c\ll T_2^s)		\label{eqnCritFast}\\
 &N\gg1,	\hspace{13mm}(\mbox{free-induction decay},\;\tau_c\gg T_2^s)	\label{eqnCritSlowSt}\\
 &N\gg\tau_c/T_2^\mrm e.	\hspace{21.5mm}(\mbox{Hahn echo},\;\tau_c\gg T_2^s)\label{eqnCritSlowE}
\end{align}
To obtain Eq.~\eq{eqnCritFast}, we have used Eqs.~\eq{eqnTdM} and \eq{eqnTdNMst} to express $T_\mrm{2,sl.}^\ast$ in terms of $\tau_c$ and $\TdM$, replacing again $\tau_n\rightarrow\tau_c$. Similarly, to obtain Eq.~\eq{eqnCritSlowE}, we have used Eqs.~\eq{eqnTdNMst} and \eq{eqnTdNMe} to express $T_\mrm{2,sl.}^\ast$ in terms of $\tau_c$ and $T_2^\mrm e$ in the slow-noise limit.

Eq.~\eq{eqnCritSlowE} shows that the minimum number of fluctuators required for the leading term $f_2(t)$ to dominate over the subleading term can become arbitrarily large in the limit $\tau_c/\Tde\rightarrow\infty$, corresponding to fluctuators with a vanishing equilibration rate. This result is consistent with the results of Ref.~\onlinecite{galperin2007non}, in which the authors showed that the Hahn-echo coherence factor for a qubit coupled to a two-level fluctuator with a switching rate $1/\tau_n\ll\D\xi_n$ shows a strong non-Gaussian behavior. Non-Gaussian corrections to the qubit coherence factor have also been considered in Ref.~\onlinecite{cywinski2008how} for various dynamical decoupling sequences.

When the criteria given by Eqs.~\eq{eqnCritFast} to \eq{eqnCritSlowE} are satisfied, the leading contribution to $|\tilde C(t)|$ (corresponding to the theory explained in Sec.~\ref{secTwoLevel}) dominates over the subleading term.

%%%%%%%%%%%%%%%%%%%%%%%%%%%%%%%%%%%%%%%%%%%%%%%%%%%%%%%%%%%%%%%%%%%%%%%%%%%%%%%%%%%%%%%%%%%%%%%%%%%%%%%%

\section{Electron-phonon coupling strength \label{secElPhon}}

Introducing a deformation potential tensor $\Xi^\chi$ for each conduction band minimum, the deformation contribution to $A_{\mvec q\ld\chi}$ is\cite{mahan1990many,yu1996fundamentals}
\begin{equation}
 A_{\mvec q\ld\chi}^\mrm d=\frac12\sqrt{\frac{\hbar}{2\rho\ups\w_{\mvec q\ld}}}\sum_{ij}\Xi_{ij}^{\chi}(q_i \xi_{\mvec q\ld}^j+q_j \xi_{\mvec q\ld}^i),\label{eqnAqld}
\end{equation}
where $\rho$ is the mass density of the sample and $\ups$ its volume. We have also introduced $\mvec \xi_{\mvec q \ld}$, the vector indicating the propagation direction of the phonon mode $\mvec q\ld$ with angular frequency $\w_{\mvec q \ld}$. The effect of shear strains on the single conduction-band minimum of GaAs is negligible relative to the effect of volume dilations.\cite{yu1996fundamentals} The deformation-potential tensor for GaAs thus reduces to $\Xi_{ij}=\dt_{ij}a(\G_{1c})\simeq-8.6$~eV. In silicon, there are six conduction-band minima at $\mvec k$-points along the six directions equivalent to $[100]$, at roughly $85\,\%$ of the distance to the the Brillouin-zone boundary.\cite{yu1996fundamentals} We label these minima as $\pm\mvec x$, $\pm\mvec y$, and $\pm\mvec z$. Using these labels, the silicon deformation-potential tensor takes the form\cite{herring1956transport,yu1996fundamentals}
\begin{align}
 \Xi^\chi_{ij}&=\dt_{ij}\Xi_d\!\\
 &+\sum_{s=\pm}\!(\dt_{\chi,s\mvec x}\dt_{ix}\dt_{jx}\!+\!\dt_{\chi,s\mvec y}\dt_{iy}\dt_{jy}\!+\!\dt_{\chi,s\mvec z}\dt_{iz}\dt_{jz})\Xi_u, (\mrm{Si})\notag
\end{align}
where $\Xi_d\simeq5$~eV and $\Xi_u\simeq8.77$~eV.\cite{yu1996fundamentals}

Crystalline silicon is not piezoelectric since the diamond lattice has inversion symmetry. In contrast, GaAs has a zincblende structure, for which the piezoelectric contribution is\cite{mahan1972polarons}
\begin{equation}
 A_{\mvec q\ld}^\mrm p=2e\sqrt{\frac{\hbar}{2\rho\ups\w_{\mvec q\ld}}}\frac{e_{14}}{\veps}\frac{q_xq_y\xi^z_{\mvec q\ld}+q_yq_z\xi^x_{\mvec q\ld}+q_zq_x\xi^y_{\mvec q\ld}}{q^2},
\end{equation}
where $e$ is the elementary charge, $e_{14}$ is the 14 element of the piezoelectric tensor in Voigt notation, and $\veps$ is the static dielectric constant of the material.

%%%%%%%%%%%%%%%%%%%%%%%%%%%%%%%%%%%%%%%%%%%%%%%%%%%%%%%%%%%%%%%%%%%%%%%%%%%%%%%%%%%%%%%%%%%%%%%%%%%%%%%%

\section{Fluctuator equilibration rate for the phonon sum process \label{secSum}}

The fluctuator equilibration rate for the phonon sum process is
\begin{align}
 \frac{1}{\tau_n^\Sg}\simeq
      \left[\frac{\Xi^4}{560}\frac{\wn^9}{\vLA^8}
      + \frac{9}{1225}\left(\Xi\frac{ee_{14}}{\veps}\right)^2\left(1+\frac43\z^2\right)\frac{\wn^7}{\vLA^6}	\right.\notag\\
      + \left.
	\frac{27}{6125}\left(\frac{ee_{14}}{\veps}\right)^4\left(1+\frac43\z^2\right)\frac{\wn^5}{\vLA^4}
      \right]\frac{2\pi(\sq_0^n)^4 (\kB T)^2}{\hbar^4(\w_\g^n)^2 m_\mrm{at}^2\wD^6}.\label{eqnRateSum}
\end{align}
The symbols in the above equation are defined below Eqs.~\eq{eqnRate1phonon} and \eq{eqnRateRaman}.

%%%%%%%%%%%%%%%%%%%%%%%%%%%%%%%%%%%%%%%%%%%%%%%%%%%%%%%%%%%%%%%%%%%%%%%%%%%%%%%%%%%%%%%%%%%%%%%%%%%%%%%%

%\bibliography{bibliographie,articlesNotes}

\bibliography{article}

\end{document}